\documentclass[twocolumn,showpacs,superscriptaddress,preprintnumbers,prd]{revtex4-1}
\usepackage{graphicx,amsmath,amssymb,dcolumn,bm}
\usepackage{fixmath}
\usepackage{feynmp}
\usepackage{epic,eepic}
\usepackage{slashed}
\allowdisplaybreaks[2]
\usepackage{calrsfs}
\DeclareMathAlphabet{\pazocal}{OMS}{zplm}{m}{n}

\begin{document}
\preprint{KEK-TH-2258, J-PARC-TH-0227}
\title{Transverse-momentum-dependent parton distribution functions\\
up to twist 4 for spin-1 hadrons}
\author{S. Kumano}
\affiliation{KEK Theory Center,
             Institute of Particle and Nuclear Studies, \\
             High Energy Accelerator Research Organization (KEK), \\
             Oho 1-1, Tsukuba, Ibaraki, 305-0801, Japan}
\affiliation{J-PARC Branch, KEK Theory Center,
             Institute of Particle and Nuclear Studies, KEK, \\
           and Theory Group, Particle and Nuclear Physics Division, 
           J-PARC Center, \\
           Shirakata 203-1, Tokai, Ibaraki, 319-1106, Japan}
\author{Qin-Tao Song}
\email[]{songqintao@zzu.edu.cn}
\affiliation{School of Physics and Microelectronics, Zhengzhou University, \\
             Zhengzhou, Henan 450001, China}
\date{December 22, 2020}

\begin{abstract}
We show possible transverse-momentum-dependent parton distribution functions
(TMDs) for spin-1 hadrons including twist-3 and 4 functions in addition to 
the leading twist-2 ones by investigating all the possible decomposition
of a quark correlation function in the Lorentz-invariant way.
The Hermiticity and parity invariance are imposed in the decomposition;
however, the time-reversal invariance is not used due to
an active role of gauge links in the TMDs.
Therefore, there exist time-reversal-odd functions
in addition to the time-reversal even ones in the TMDs.
We list all the functions up to twist-4 level
because there were missing terms associated 
with the lightcone vector $n$ in previous works
on the twist-2 part and there was no correlation-function study
in the twist-3 and 4 parts for spin-1 hadrons.
We show that 40 TMDs exist in the tensor-polarized spin-1 hadron
in twists 2, 3, and 4. Some expressions of twist-2 structure functions 
are modified from previous derivations due to the new terms with $n$, 
and we find 30 new structure functions in twists 3 and 4 in this work.
Since time-reversal-odd terms of the collinear correlation
function should vanish after integrals over 
the partonic transverse momentum, we obtain new sum rules
for the time-reversal-odd structure functions,
$ \int d^2 k_T g_{LT} 
 = \int d^2 k_T h_{LL} = \int d^2 k_T h_{3LL} =0$.
In addition, we indicate that 
new transverse-momentum-dependent fragmentation functions
exist in tensor-polarized spin-1 hadrons.
The TMDs are rare observables to find explicit color degrees of freedom
in terms of color flow, which cannot be usually measured because the color 
is confined in hadrons. 
Furthermore, the studies of TMDs enable us not only to find three-dimensional 
structure of hadrons, namely hadron tomography including transverse structure,
but also to provide unique opportunities for creating interesting 
interdisciplinary physics fields such as gluon condensates, 
color Aharonov-Bohm effect, and color entanglement.
The tensor structure functions
may not be easily measured in experiments.
However, high-intensity facility such as 
the Thomas Jefferson National Accelerator Facility (JLab),
the Fermilab Main Injector, and 
future accelerators like electron-ion collider (EIC) may probe 
such observables. 
In addition, since the Nuclotron-based Ion Collider fAcility (NICA)
focuses on spin-1 deuteron structure functions, there is
a possibility to study the details of polarized structure 
functions of the deuteron at this facility.
\end{abstract}
\maketitle

\section{Introduction}
\label{sec:1}

It had been taken for granted that the proton spin consists
of three quark spins in the naive quark model. 
However, the European Muon Collaboration experiment found 
that the quark contribution accounts for only 20$-$30\% of 
the proton spin \cite{emc-1988}, and the rest 
should be carried by gluon-spin and partonic 
orbital-angular-momentum (OAM) contributions \cite{nucleon-spin}.
In order to figure out the partonic OAM contributions, 
one needs to study three-dimensional structure functions which include 
generalized parton distributions (GPDs) \cite{gpds}, 
generalized distribution amplitudes 
(GDAs or timelike GPDs) \cite{gdas}, 
and transverse-momentum-dependent parton distributions (TMDs) \cite{tmds}. 

The TMDs indicate the parton distributions as the function of
the partonic transverse momentum $k_T$ in addition to
the longitudinal momentum fraction $x$.
The color flow appears explicitly in the TMDs, although
it does not show up easily in other observables because of
the color confinement.
They have interesting application to other fields of physics,
such as the gluon condensate \cite{gluon-condensate}, 
color Aharonov-Bohm effect \cite{AB-effect,qwork-mulders}, 
and color entanglement \cite{color-entanglement}.
The TMD studies are new tools to investigate
the color degrees of freedom and to explore new interdisciplinary
fields beyond standard hadron physics.

The TMDs have been investigated for the spin-1/2 proton;
however, spin-1 hadrons and nuclei such as the deuteron have
new interesting aspects due to the spin-1 nature.
In the charged lepton scattering, there are four collinear structure functions
$b_{1-4}$ in the deuteron in addition to the ones for the nucleon
\cite{fs83,Hoodbhoy:1988am}.
Among them, the leading-twist functions are $b_1$ and $b_2$, which 
are related with each other by the Callan-Gross type relation 
$2x b_1 = b_2$ in the scaling limit $Q^2 \to \infty$.
These structure functions are expressed by tensor-polarized
parton distribution functions (PDFs).
In addition, there is the gluon transversity distribution 
\cite{jlab-gluon-trans} in the leading twist.
For $b_1$, there were measurements by the HERMES 
Collaboration \cite{Airapetian:2005cb}; however,
the magnitude and $x$ dependence of $b_1$ are very different
from conventional convolution calculations based
on a standard deuteron model with $D$-state admixture
\cite{b1-convolution,tagged-ed}.
Furthermore, the HERMES data indicated a finite sum 
$\int dx b_1 (x) 
  = [ \, 0.35 \pm 0.10 \, (\text{stat}) \pm 0.18 \, (\text{sys}) \, ]$
\cite{Airapetian:2005cb},
which indicates a finite tensor-polarized antiquark 
distribution according to the parton-model sum rule \cite{b1-sum}
$ \int dx \, b_1 (x)
    = - \lim_{t \to 0} \frac{5}{24} \, t \, F_Q (t) 
     + \sum_i e_i^2 \int dx \, \delta_{_T} \bar q_i (x) $,
where $F_Q (t)$ is the electric quadrupole form factor of the hadron,
and $\delta_{_T} \bar q_i$ is the tensor-polarized antiquark 
distribution.
The first term vanishes, so that a finite sum of $b_1$ indicates
a finite tensor-polarized antiquark distribution.
The vanishing first term comes from the fact that 
the valence-quark number does not depend on the tensor polarization,
whereas it depends on the flavor in the Gottfried sum (1/3)
\cite{gottfried}.

Since the finite $b_1$ sum indicates a new topic on tensor-polarized
antiquark distributions and the standard convolution-model
distribution for $b_1$ is very different from the HERMES data
\cite{b1-convolution},
a new high-energy spin physics field could be foreseen by
investigating tensor-polarized structure functions.
Experimentally, there is an experimental proposal to measure 
$b_1$ accurately measured at 
at Thomas Jefferson National Accelerator Facility (JLab) \cite{jlab-b1} 
and tensor-polarized PDFs could be measured at Fermilab
by the SpinQuest (E1039) experiment 
\cite{Fermilab-dy} by the proton-deuteron
Drell-Yan process with the tensor-polarized deuteron target
\cite{Keller:2020wan}.
The proton-deuteron Drell-Yan formalism was given 
in Ref.~\cite{pd-drell-yan},
and tensor-polarized spin asymmetries were estimated in 
Ref.~\cite{Kumano:2016ude} based on the parametrization
for the tensor-polarized PDFs of Ref.~\cite{tensor-pdfs}.
There were also GPD studies on the spin-1 deuteron
and $\rho$ meson \cite{trans-gpds}
and fragmentation-function studies on spin-1 hadrons \cite{Ji-1994}.

On the gluon transversity, there is an experimental plan
to measure it at JLab \cite{jlab-gluon-trans}
and there is a possibility to study it at Fermilab
by using the proton-deuteron Drell-Yan process with
the linearly polarized deuteron \cite{Kumano:2020gfk}.
Furthermore, there are possibilities at 
NICA (Nuclotron-based Ion Collider fAcility) \cite{nica}
and GSI-FAIR (Gesellschaft f\"ur Schwerionenforschung-Facility for 
Antiproton and Ion Research).
Since the spin-1/2 proton and neutron in the deuteron cannot
contribute to the gluon transversity, it is an appropriate quantity
to find new hadron physics beyond the simple bound system of 
the nucleons.

These $b_1$ and gluon transversity distribution are collinear
functions as the function of $x$.
In this work, we investigate possible TMDs for spin-1 hadrons
especially by considering the tensor polarization.
The TMDs are generally defined from the quark correlation function.
The quark correlation function and its relations to PDFs 
were investigated for the spin-1/2 nucleon in 
Refs.\,\cite{Ralston:1979ys,tangerman-prd-1995},
and additional terms were studied in Ref.\,\cite{pd-drell-yan}
for the spin-1 deuteron.
The quark TMD correlation function was decomposed into possible terms
by considering Lorentz invariance, Hermiticity, and parity conservation
in Refs.~\cite{ Mulders:1995dh, tangerman-th} 
for spin-1/2 proton, and then the TMDs were introduced by integrating 
the correlation function over the minus component 
of the quark lightcone momentum.

Much progress has been made in the TMD studies based on these works;
however, it was found later that 
the decomposition of the quark correlation function in 
Refs.~\cite{Ralston:1979ys, Mulders:1995dh, tangerman-th, pd-drell-yan} 
was not complete. 
The quark correlation function depends on the lightcone vector $n$,
which is defined in Eq.\,(\ref{eqn:lightcone-n-nbar}),
due to the gauge link or the Wilson line $W(0, \xi | n)$ 
which guarantees the color gauge invariance of the correlation function
\cite{Goeke:2003az, Bacchetta:2004zf, Goeke:2005hb}
as defined later in Eqs.\,(\ref{eqn:wilson-1}), 
(\ref{eqn:wilson-2}), and (\ref{eqn:wilson-4}).
The vector $n$ specifies the direction along the gauge link.
The complete decomposition of the quark correlation function
was made by introducing twenty new terms which are associated
with the lightcone vector $n$ 
for the spin-1/2 nucleon in Ref.~\cite{Goeke:2005hb}. 
Even though these new terms in the correlation function
do not give rise to new TMDs at the leading-twist level, 
they bring new observables in the semi-inclusive 
deep inelastic scattering (SIDIS) 
which are expressed by
the new twist-3 TMDs \cite{Bacchetta:2004zf}.
The new terms in the correlation function 
also affect relations of the collinear PDFs. 
For example, several Lorentz invariance relations  
for the PDFs were obtained \cite{Boer:1997nt}
based on the decomposition of the quark
correlation function in Refs.~\cite{ Mulders:1995dh, tangerman-th}, 
and these relations were modified if one considered
the complete decomposition of the correlation function
\cite{Kundu:2001pk, Goeke:2003az}. Moreover, the Wandzura-Wilczek
relation \cite{Wandzura:1977qf} was reinvestigated in 
Refs.~\cite{Metz:2008ib, Accardi:2009au}, 
it was found that the Wandzura-Wilczek relation is not satisfied
due to another new twist-3 term.

These additional terms due to $n$ were studied for the spin-1/2 nucleon
\cite{Goeke:2005hb}. The purpose of this work is to derive new TMDs
associated with $n$ for spin-1 hadrons up to the twist-4 level. 
As for a stable spin-1 hadron or nucleus for experiments,
the deuteron is the most simple and stable particle.
It is known that there are additional structure functions 
in the spin-1 deuteron in comparison with the spin-1/2 nucleon, 
since both vector polarization and tensor polarization are available
in the deuteron. 
The tensor polarization does not exist for the spin-1/2 nucleon, 
and it could be used to investigate new aspects in the deuteron.
The deuteron is a weakly bound state of proton and neutron. 
However, the spin-1/2 proton and neutron do not contribute 
directly to the tensor structure,
which is an interesting aspect in studying the deuteron's
tensor polarizations.

As for theoretical studies, the spin-1 TMDs were investigated in 
Refs.~\cite{pd-drell-yan,bate2000,mulders2001} 
and T-even TMDs were calculated 
in an effective hadron model for the $\rho$ meson
\cite{ninomiya-spin-1}.
However, the terms associated with the lightcone vector $n$ were not 
included in the  decomposition of the quark-quark correlation function in 
Ref.\,\cite{bate2000}.
Since these new terms could have a significant impact 
on the structure-function studies in the spin-1 hadrons,
we show the complete decomposition of the quark correlation 
function for the spin-1 hadrons in this paper.
In this paper, the transverse-momentum-dependent 
quark correlation function and parton distribution functions
are explained in Sec.\,\ref{sec:2}.
Next, possible TMDs are obtained by decomposing the quark
correlation function in Sec.\,\ref{sec:3}.
Then, our studies are summarized in Sec.\,\ref{sec:4}.

\section{Transverse-momentum-dependent parton distribution functions}
\label{sec:2}

In this section, we introduce the TMDs and discuss motivations
for investigating the TMDs.
First, the three-dimensional structure functions are explained
as a field of hadron tomography from generalized TMDs and Wigner
functions as generating functions in Sec.\,\ref{3d}.
The quark correlation function is introduced with
proper gauge links, which play an important role in the TMD physics
in Sec.\,\ref{correlation-fun-link}.
We show that the color flows, expressed by the gauge links, are 
different in the SIDIS
and DY processes. This fact leads to the sign
change in the time-reversal-odd quark TMDs.
The time-reversal properties of the quark correlation function
are discussed in Sec.\,\ref{time-reversal}.

\subsection{Hadron tomography by three-dimensional structure functions}
\label{3d}

\begin{figure}[t]
 \vspace{+0.15cm}
\begin{center}
   \includegraphics[width=8.5cm]{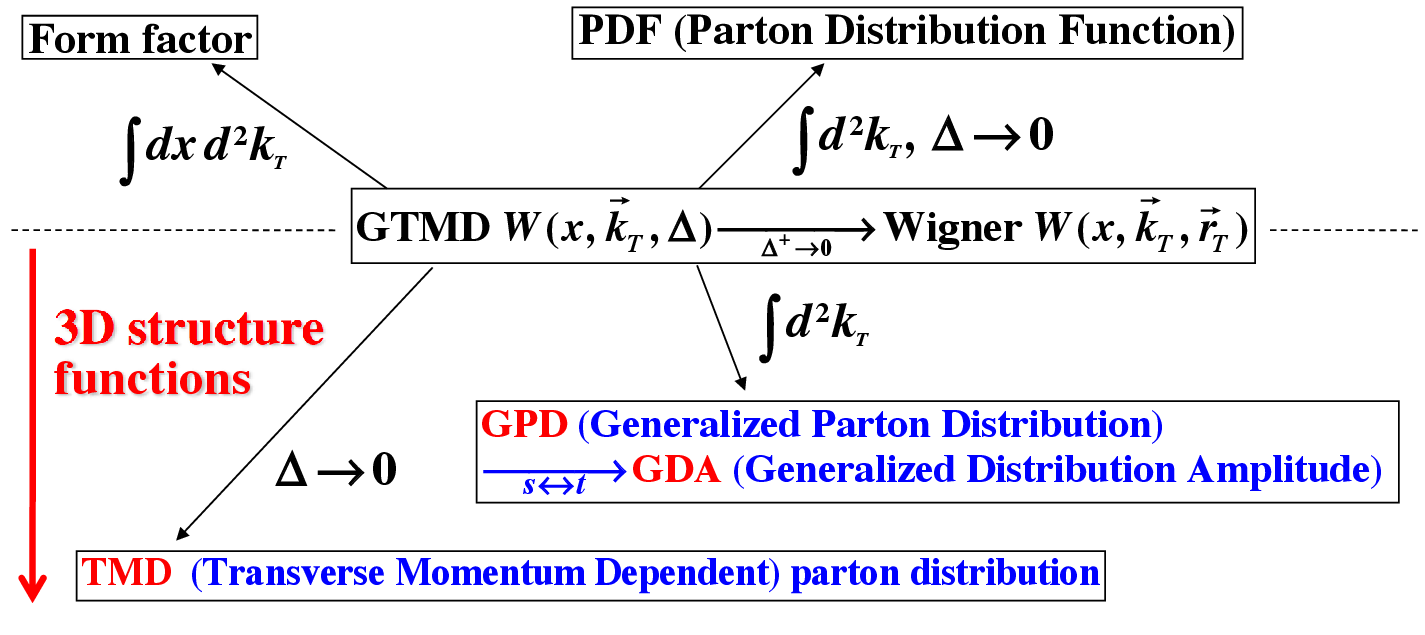}
\end{center}
\vspace{-0.6cm}
\caption{
Three dimensional structure functions (TMD, GPD, GDA) 
from the generalized transverse-momentum-dependent 
parton distribution (GTMD) and the Wigner function, 
together with the form factor and parton distribution function.
}
\label{fig:3d-structure}
\vspace{-0.30cm}
\end{figure}

Until recently, hadron structure had been investigated by
electromagnetic form factors and parton distribution functions (PDFs).
However, recent studies focus on 3D aspects
including the transverse structure in addition to 
the longitudinal one along the hadron-momentum direction.
The 3D structure studies were originally motivated
for finding the origin of nucleon spin including
the partonic orbital-angular momenta (OAM). 
The OAM contribution to the nucleon spin should be probed by
one of 3D structure functions, especially the GPDs.
However, the hadron tomography, namely the 3D structure of hadrons,
has deeper meaning in the sense that it could probe
gravitational form factors of hadrons without relying on
explicit graviton interactions \cite{gdas}.
The hadron tomography has been investigated by three types 
of structure functions, TMDs, GPDs, and GDAs (or timelike GPDs). 
They are obtained from the generating functions called 
generalized transverse-momentum-dependent 
parton distributions (GTMDs) and the Wigner functions as illustrated
in Fig.\,\ref{fig:3d-structure}.
The TMDs are obtained by taking the forward limit $\Delta \to 0$,
where $\Delta$ is the momentum transfer from the initial hadron to
the final one ($\Delta = P' - P$),
and the GPDs are obtained by integrating the GTMDs
over the parton's transverse momentum $\vec k_T$.
The GDAs are related to the GPDs by the $s$-$t$ crossing, where
$s$ and $t$ are Mandelstam variables.

\subsection{Quark correlation functions and color flow}
\label{correlation-fun-link}

\begin{figure}[b]
 \vspace{-0.30cm}
\begin{center}
   \includegraphics[width=5.5cm]{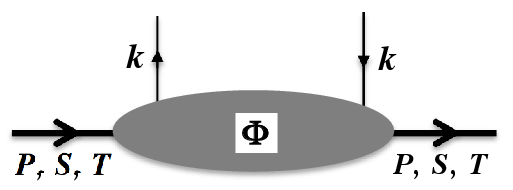}
\end{center}
\vspace{-0.6cm}
\caption{
Quark correlation function $\Phi$ with the quark (hadron)
momentum $k$ ($P$) and the hadron vector
and tensor polarizations $S$ and $T$, respectively,}
\label{fig:correlation-q}
\vspace{-0.00cm}
\end{figure}

The TMDs and collinear PDFs are defined from the quark correlation function
\begin{align}
& \Phi_{ij}^{[c]} (k, P, S, T)  =
\int  \! \frac{d^4 \xi}{(2\pi)^4} \, e^{ i k \cdot \xi}
\nonumber \\[-0.10cm]
& \ \ \ \ \ \ \ \ \ \ 
\times
\langle \, P , S, T \left | \, 
\bar\psi _j (0) \,  W^{[c]} (0, \xi)  
 \psi _i (\xi)  \, \right | P, \,  S, T\rangle ,
\label{eqn:correlation-q}
\end{align} 
which is illustrated in Fig.\,\ref{fig:correlation-q}.
It may be denoted as $\Phi_{q/H,\,ij}^{[c]}$ with 
$q=u$,\,$d$,\,$s$,\,$\cdots$, but we abbreviated 
the notations on the quark flavor $q$ and the hadron $H$.
The correlation function is related to
the amplitude to extract a parton 
from a hadron and then to insert it into the hadron
at a different spacetime point $\xi$.
Here, $\psi$ is the quark field, 
$\xi$ is a four-dimensional space-time coordinate,
$k$ and $P$ are the quark and hadron momenta,
$S$ and $T$ are vector and tensor polarizations of the hadron,
and $W^{[c]}(0, \xi)$ is called the gauge link or the Wilson line
so as to satisfy the color gauge invariance.
It is defined by the path-ordered exponential ($\pazocal{P}$)
\begin{align}
W^{[c]}(0, \xi) = \pazocal{P}  \exp \left [
                 -i \, g \int_{0,\,c}^{\, \xi} 
               d\xi \cdot A (\xi) \right ] .
\label{eqn:wilson-1}
\end{align}
The gauge link indicates gluon interactions between quarks 
for satisfying the gauge invariance.
Here, $c$ indicates the integral path, and $A_\mu (\xi)$ is
$A_\mu (\xi) = A_\mu ^a (\xi) \, t^a$ by including 
the SU(3) generator $t^a$ expressed by the Gell-Mann matrix $\lambda^a$
as $t^a = \lambda^a /2$ with the color index $a$.
The antiquark correlation function is defined in the same way
\cite{Kumano:2020gfk}.

The TMDs and collinear PDFs for quarks are then given by the quark 
correlation functions as \cite{Kumano:2020gfk}
\begin{align}
\Phi^{[c]} (x, k_T, P, S, T ) & = \! \int \! dk^+ dk^- \, 
               \Phi^{[c]} (k, P, S, T \, |n) 
\nonumber \\[-0.20cm]
& \ \hspace{1.8cm} \
 \times \delta (k^+ \! -x P^+) ,
\nonumber \\[-0.10cm]
\Phi (x, P, S, T ) & 
  = \! \int \! d^2 k_T \, \Phi^{[c]} (x, k_T, P, S, T ) ,
\label{eqn:correlation-tmd}
\end{align}
where $k_T$ is the quark transverse momentum, 
$\Phi^{[c]} (x, k_T, P, S, T )$ is
the transverse-momentum-dependent correlation function
which is related later to the TMDs, and $\Phi (x, P, S, T)$ is
the collinear correlation function.
The lightcone $\pm$ momenta are defined by 
$a^\pm = (a^0 \pm a^3)/\sqrt{2}$, and
the lightcone vectors $n$ and $\bar n$ are given by 
\begin{align}
n^\mu =\frac{1}{\sqrt{2}} (\, 1,\, 0,\, 0,\,  -1 \, ), \ \ 
\bar n^\mu =\frac{1}{\sqrt{2}} (\, 1,\, 0,\, 0,\,  1 \, ) .
\label{eqn:lightcone-n-nbar}
\end{align} 
The integral path depends on the lightcone direction $n^-$,
which is explicitly shown
as the $n$ dependence in
Eq.\,(\ref{eqn:correlation-tmd}).
We note that there is no link-path dependence $c$
in the collinear correlation function $\Phi (x, P, S, T)$
as shown in this section.
From Eqs.\,(\ref{eqn:correlation-q}) and (\ref{eqn:correlation-tmd}),
the transverse-momentum-dependent correlation function is
expressed as
\begin{align}
& \Phi_{ij}^{[c]} (x, k_T, P, S, T )  =
\int  \! \frac{d\xi^- d^2 \xi_T}{(2\pi)^3} \,
 e^{ik^+ \xi^- -i \vec k_T \cdot \vec \xi_T}
\nonumber \\[-0.05cm]
& \ 
\times
\langle \, P , S, T \left | \, 
\bar\psi _j (0) \,  W^{[c]} (0, \xi \, | n)  
 \psi _i (\xi)  \, \right | P, \,  S, T \, \rangle _{\xi^+ =0} ,
\label{eqn:correlation-q-tmd}
\end{align} 
with the plus lightcone momentum $k^+ = x P^+$
by taking the hadron momentum direction as the third axis.
 
\begin{figure}[b]
 \vspace{-0.30cm}
\begin{center}
   \includegraphics[width=5.5cm]{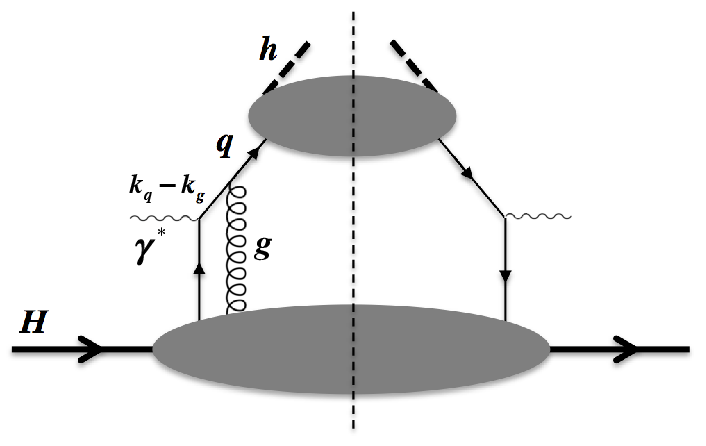}
\end{center}
\vspace{-0.7cm}
\caption{A typical semi-inclusive DIS process 
$\gamma^* + H \to h +X$ ($\ell \to \gamma^* \ell'$, $\ell=e$ or $\mu$)
with a gluon interaction in the final state.}
\label{fig:sidis-ex1}
\vspace{-0.00cm}
\end{figure}

\begin{figure}[t]
 \vspace{-0.30cm}
\begin{center}
   \includegraphics[width=3.2cm]{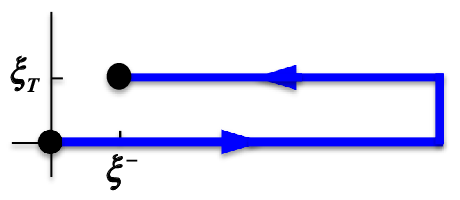}
   \ \ \ \ \ \ \ \ 
   \includegraphics[width=4.0cm]{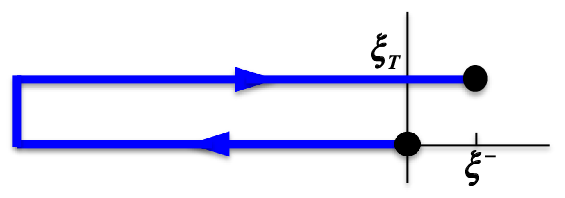} \\
\hspace{-0.5cm}\vspace{-0.40cm}
{\boldmath\normalsize$(a)$}
\hspace{+3.5cm}
{\boldmath\normalsize$(b)$}
\end{center}
\vspace{-0.1cm}
\caption{Gauge link for $(a)$ semi-inclusive DIS 
with the spacelike correlation function $\Phi^{[+]}$
and $(b)$ Drell-Yan process
with the timelike correlation function $\Phi^{[-]}$.}
\label{fig:gauge-link-sidis-dy}
\vspace{-0.00cm}
\end{figure}

The details of the gauge link for the 
SIDIS are, for example, 
explained in Ref.~\cite{Boer-2003}.
Resummations of processes with intermediate gluons, as 
typically shown in Fig.\,\ref{fig:sidis-ex1}, lead to
the gauge link 
\cite{tmds,Collins-1982-2002,Ji-2002,Belitsky-2003,Boer-2003}.
Here, the gauge link $W^{[c]} (0, \xi |n)$ for the TMD correlation function
in the SIDIS process ($c=+$) is given by
\begin{align}
W^{[+]} (0, \xi \, | n) 
    & 
=         [\, 0, \vec 0_T;\, \infty, \vec 0_T \, ] \,
          [\, \infty, \vec 0_T;\, \infty, \vec{\xi}_T \, ]
\nonumber \\
    & \ \ \ \ \ \ 
     \times 
        [\, \infty, \vec{\xi}_T;\,  \xi^-,  \vec{\xi}_T \,]_{\xi^+ =0} .
\label{eqn:wilson-2}
\end{align}
Here, the notation 
$[\, a^-, \vec a_T;\, b^-, \vec b_T \, ]$
(or doted as $W (a,b \, |n)$)
indicates the gauge line connecting 
$a = (a^+=0, a^-, \vec a_T)$
to $b = (b^+ =0, b^-, \vec b_T)$ 
along the straight lightcone direction of $\xi^-$ 
(namely, plus direction of $n^-$),
and $[\, a^-, \vec a_T;\, b^-, \vec b_T \, ]$
($W (a,b \, | \, \vec\xi_T)$)
is the link along the transverse direction $\vec\xi_T$:
\begin{align}
W (a,b \, |n) & = [\, a^-, \vec a_T;\, b^- , 
                   \vec b_T \,(= \vec a_T) \, ] 
\nonumber \\[-0.05cm]
& \! \! \! \! \! \! \! \!  
\equiv \mathcal{P} \exp \left [ -ig \int_{a^-}^{\, b^-} 
   d\xi^-  A^+ (\xi) \right ]
   _{\substack{\xi^+ =a^+ = b^+\\ \!\!\vec\xi_T =\vec a_T=\vec b_T}} ,
\nonumber \\[-0.10cm]
W (a,b \, | \, \vec\xi_T) & = 
[\, a^-, \vec a_T;\, b^- \,(= a^-), \vec b_T \, ] 
\nonumber \\[-0.05cm]
& \! \! \! \! \! \! \! \!  
\equiv \mathcal{P} \exp \left [
                 -ig \int_{\vec a_T}^{\, \vec b_T} \! \! \!
              d \vec \xi_T \cdot \vec A_T (\xi) \right ] 
   _{\xi^\pm =a^\pm = b^\pm} .
\label{eqn:wilson-3}
\end{align}
The superscript $[+]$ 
of $W^{[+]}$ in Eq.\,(\ref{eqn:wilson-2})
indicates the integral path
along the plus direction in the coordinate $\xi^-$
in the first link step.
The final expression for the link path of Eq.\,(\ref{eqn:wilson-2})
is shown in $(a)$ of Fig.\,\ref{fig:gauge-link-sidis-dy}.
The path $c=+$ consists of the three gauge links.
The path dependence of the gauge link is important 
in TMD physics, as we show the difference between the TMDs
of the SIDIS and the Drell-Yan process in the following.

\begin{figure}[b]
 \vspace{-0.00cm}
\begin{center}
   \includegraphics[width=5.5cm]{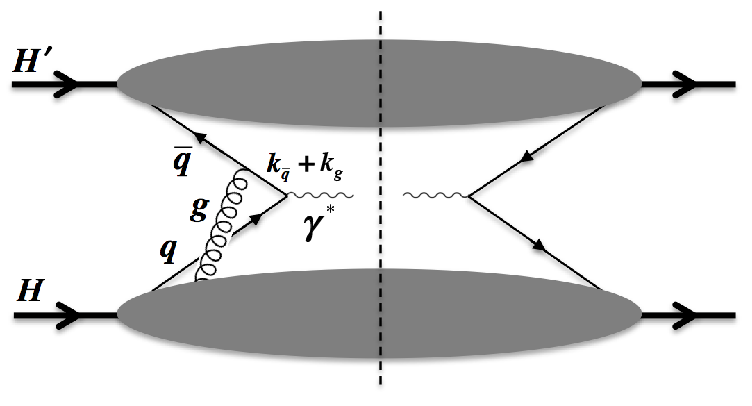}
\end{center}
\vspace{-0.7cm}
\caption{A typical Drell-Yan process 
$H + H' \to \gamma^* +X$ ($\gamma^* \to \mu^- \mu^+$)
with a gluon interaction in the initial state.}
\label{fig:dy-ex1}
\vspace{-0.30cm}
\end{figure}

A typical Drell-Yan process 
$H_1 + H_2 \to \gamma^* +X$ ($\gamma^* \to \mu^- \mu^+$)
with an intermediate gluon is shown in Fig.\,\ref{fig:dy-ex1}.
We note that the gluon exchange occurs in the initial state,
whereas it does in the final state in the SIDIS 
\cite{brodsky-2002}
as shown in Fig.\,\ref{fig:sidis-ex1}. It leads to the path difference
in the gauge link and subsequently in the sign difference
in both TMDs.
The cross sections of these SIDIS and Drell-Yan processes 
are calculated \cite{tmds,Boer-2003,Belitsky-2003},
and it was found that 
the color flows in the opposite lightcone direction
between the SIDIS and Drell-Yan processes.
Therefore, the gauge link for the Drell-Yan process ($c=-$) 
is given by
\begin{align}
W^{[-]} (0, \xi \, | n) &  
=         [\, 0, \vec 0_T;\, -\infty, \vec 0_T \, ] \,
          [\, -\infty, \vec 0_T;\, -\infty, \vec{\xi}_T \, ]
\nonumber \\
    & \ \ \ \ \ \ 
     \times 
        [\, -\infty, \vec{\xi}_T;\,  \xi^-, 
           \vec{\xi}_T \,]_{\xi^+ =0} ,
\label{eqn:wilson-4}
\end{align}
as shown in Fig.\,\ref{fig:gauge-link-sidis-dy}$(b)$.
We notice that the gauge links of the SIDIS and Drell-Yan are opposite
in the $\xi^-$ direction, which results in the sign change
in the time-reversal-odd TMDs as shown in Eq.\,(\ref{eqn:tmd-sign}).
The superscript $[-]$ indicates the integral path of the first link step
along the minus direction in the coordinate $\xi^-$.

If the transverse momentum is integrated as in Eq.\,(\ref{eqn:correlation-tmd}),
the collinear correlation function of Eq.\,(\ref{eqn:correlation-q-tmd}) 
becomes
\begin{align}
& \! \! \! \! 
\Phi_{ij} (x, P, S, T )  =
\int  \! \frac{d\xi^-}{2\pi} \,
 e^{ixP^+ \xi^-}
\nonumber \\
& \! \! \! \! \! \! 
\times \! 
\langle \, P , S, T \left | \, 
\bar\psi _j (0) \,  W (0, \xi \, |n)  
 \psi _i (\xi)  \, \right | \! P, \,  S, T\rangle _{\xi^+ =0, \, \vec\xi_T=0} ,
\label{eqn:correlation-q-tmd-2}
\end{align} 
where $k^+ = x P^+$ is used.
Since the gauge link is the straight line from $\xi^- =0$
to $\xi^-$ with $\xi^+ =0$ and  $\vec\xi_T=0$,
these collinear functions are not dependent on the gauge-link path 
like the TMDs and a process-dependent relation like Eq.\,(\ref{eqn:tmd-sign})
does not exist. The color flow can be probed 
only if transverse-momentum-dependent observables are investigated.

\subsection{Existence of time-reversal-odd structure functions
and their color-flow dependence}
\label{time-reversal}

Here, we show properties of the TMD correlation function
under the time reversal for defining time reversal even and
odd functions. However, one should note that the existence of
the time-reversal-odd functions does not mean the violation
of the time-reversal invariance as explained in this subsection.

\begin{table*}[t]
\scriptsize
\renewcommand{\arraystretch}{2.0}
\begin{tabular}{|c|c|c|c|} \hline
Quantity      &   Hermite   &  Parity      &  Time reversal \\ \hline
$P^\mu$       &       \         & $\bar P^\mu$ & $\bar P^\mu$ \\ \hline
$S^\mu$       &       \         & $- \bar S^\mu$ & $\bar S^\mu$ \\ \hline
$T^{\mu\nu}$ &       \         & $ \bar T^{\mu\nu}$ & $\bar T^{\mu\nu}$ \\ \hline
$\psi (\xi)$  
&     \             
& ${\cal P} \psi (\xi) {\cal P}^\dagger = \gamma^0 \psi (\bar\xi \,)$
& ${\cal T} \psi (\xi) {\cal T}^\dagger = (-i\gamma_5 C) 
         \psi (-\bar\xi \,)$ 
\\ \hline 
$A_\mu (\xi)$  
& $A_\mu^\dagger (\xi) = A_\mu (\xi)$             
& ${\cal P} A_\mu (\xi) {\cal P}^\dagger =  \bar A_\mu (\bar\xi)$
& ${\cal T} A_\mu (\xi) {\cal T}^\dagger =  \bar A_\mu (-\bar\xi)$ 
\\ \hline
$W (a,b)$  
& $W^\dagger (a,b) = W (b,a)$             
& ${\cal P} W (a,b) {\cal P}^\dagger =  W (\bar a,\bar b)$
& ${\cal T} W (a,b) {\cal T}^\dagger =  W (-\bar a,-\bar b)$ 
\\ \hline
$\Phi (k,P,S,T \, | n)$
& $\Phi ^ \dagger (k,P,S,T \, | n) = \gamma^0 \, \Phi (k,P,S,T \, | n) \gamma^0$             
& $\Phi (k,P,S,T \, | n) 
      = \gamma^0 \, \Phi (\bar k, \bar P, -\bar S, \bar T \, | \bar n) \gamma^0$
& $\Phi ^* (k,P,S,T \, | n) 
    = (-i\gamma_5 C) \, \Phi (\bar k,\bar P,\bar S, \bar T \, | \bar n) (-i\gamma_5 C)$ 
\\ \hline
$\Phi ^{[\pm]} (x, k_T)$  
& $\Phi ^{[\pm] \, \dagger} (x, k_T) 
      = \gamma^0 \, \Phi^{[\pm]} (x, k_T) \gamma^0 $             
& $\Phi^{[\pm]} (x, k_T) 
      = \gamma^0 \, \Phi^{[\pm]} (x, \bar k_T ) \gamma^0$
& $\Phi ^{[\pm] \, *} (x, k_T)  
    = (-i\gamma_5 C) \, \Phi ^{[\mp]} (x, \bar k_T ) (-i\gamma_5 C) $ 
\\  \hline
\end{tabular}
\caption{Properties under Hermite, parity and time-reversal transformations.
The spin $S$, tensor $T$, and lightcone vector $n$ are abbreviated 
in $\Phi ^{[\pm]} (x, k_T)$ for simplicity because their transformations 
are shown in $\Phi (k,P,S,T \, | n)$.
The charge conjugation is $C= i \gamma^2 \gamma^0$ so that 
the time-reversal factor is 
${\pazocal T}= -i\gamma_5 C = i \gamma^1 \gamma^3$. 
The time-reversal invariance condition is not imposed 
for the correction functions due to the gauge link; however,
we show the time-reversal properties in this table
to understand the T-even and odd properties in our formalism.
}
\label{table:H-P-T}
\vspace{-0.2cm}
\end{table*}
\normalsize

The parity and time reversal mean the transformations of
the space-time coordinate as
\begin{alignat}{2}
\text{Parity:} \ & 
     x^\mu = (t, \vec x\, ) \to & 
      & \ (t, - \vec x\, ) ,
\nonumber \\
\text{Time reversal:} \ & 
     x^\mu = (t, \vec x\, ) \to & 
      & \ (-t, \vec x\, ) .
\label{eqn:P-T}
\end{alignat}
The parity ($\cal P$) is a unitary operator 
and the time-reversal ($\cal T$) is an antiunitary one
\cite{leader-book,winberg-book}. The antiunitary means antilinear with
the unitarity. 
Namely, it satisfies the relations
\begin{align}
\text{Antilinear:} \, & 
    {\cal T} (a | A \rangle + b | B \rangle )
       = a^* {\cal T} | A \rangle + b^* {\cal T} | B \rangle  ,
\nonumber \\
\text{Hermite conjugate:} \, & 
    \langle A | {\cal T}^{\dagger} | B \rangle
          = \langle {\cal T} A | B \rangle ^* ,
\label{eqn:anti-linear-T}
\end{align}
where the definition of the Hermite conjugate is different
from the usual definition 
$\langle A | {\cal O}^{\dagger} | B \rangle
       = \langle {\cal O} A | B \rangle $
for the linear operator ${\cal O}$.
The momentum ($P$), spin ($S$), and tensor ($T$) 
transform under parity and time-reversal transformations as
shown in Table \ref{table:H-P-T},
where $\bar P^\mu$ and $\bar T^{\mu\nu}$ are defined by
\begin{align}
\bar P^\mu \equiv (\, P^0, - \vec P \,) = g^{\mu\alpha} P_\alpha , \ \ 
\bar T^{\mu\nu} = g^{\mu\alpha} g^{\nu\beta} T_{\alpha\beta} .
\label{eqn:Pbar-Tbar}
\end{align}
Under the parity and time reversal, the transformations 
of the quark field $\psi (\xi$) and the gluon field $A_\mu (\xi)$
\cite{bd-book-II,Itzeykson-Zuber,Boer-2003}
are shown in Table \ref{table:H-P-T}, together with 
the Hermite conjugation for $A_\mu (\xi)$.
Here, the charge conjugation $C$ is given by $C= i \gamma^2 \gamma^0$
so that the overall factor is 
${\pazocal T} = -i\gamma_5 C = i \gamma^1 \gamma^3$. 

From the transformations of the gluon field, the gauge link
$W (a,b)$ should have the transformations in 
Table \ref{table:H-P-T}.
These relations mean that the link paths are changed
due to the space-time coordinate changes $a,\, b \to \bar a,\, \bar b$ 
(or $- \bar a,\, - \bar b$) \cite{br-book,Boer-2003}.
The $\gamma^0$ factors, for example, in the Hermiticity relation 
are obtained simply by taking the Hermite conjugate of 
the correlation function in Eq.\,(\ref{eqn:correlation-q}).
The transformations for the TMD correlation function
$\Phi ^{[\pm]} (x, k_T)$ are
then given in Table \ref{table:H-P-T},
so that the time reversal transforms $\Phi ^{[+]}$ to $\Phi ^{[-]}$
and vice versa. 
The T-even and T-odd TMD functions are then defined by
\begin{align}
\Phi ^{\text{[T-even]}} (x, k_T) & = \frac{1}{2}
   \left [  \Phi^{[+]} (x, k_T)  
         +  \Phi^{[-]} (x, k_T)   \right ] ,
\nonumber \\
\Phi ^{\text{[T-odd]}} (x, k_T) & = \frac{1}{2}
   \left [  \Phi^{[+]} (x, k_T)  
         -  \Phi^{[-]} (x, k_T)   \right ] .
\label{eqn:T-even-odd-TMD-correlation}
\end{align}
If the transverse moment  $\Phi ^{[\pm] \, \mu} (x)$ is defined by
\begin{align}
\Phi ^{[\pm] \, \mu}_\partial (x) = \int d^2 k_T \, k_T^\mu \,
       \Phi ^{[\pm]} (x, k_T \,) ,
\label{eqn:trans-moment}
\end{align}
they are given by the T-odd
quark-gluon correlation function $\Phi^{[\text{T-odd}] \, \mu}_G (x,x)$ as
\cite{Boer-2003}
\begin{align}
\Phi ^{[\pm] \, \mu}_\partial (x) 
  = \Phi ^{[\text{T-even}] \, \mu} _\partial (x)
   \pm \pi \, \Phi^{[\text{T-odd}] \, \mu}_G (x,x) .
\label{eqn:quark-gluon-correlation}
\end{align}
These different link paths give rise to sign differences 
in the time-reversal-odd TMD functions.
The second term of this equation ($\Phi^{[\text{T-odd}] \, \mu}_G$)
comes from the soft gluon ($k_g^+ =0$) and it is called gluonic-pole matrix.
It suggests that the single spin asymmetries,
such as the Sivers effect, originate from this term, 
as proposed by Qiu and Sterman \cite{Qiu-1991}.
Here, the Sivers function is one of the TMDs and
it indicates the difference between a unpolarized
quark distribution in the nucleon polarized transversely 
to its momentum and the one with opposite polarization.
The T-odd TMDs exist in the single spin asymmetries in
SIDIS by the form $\Phi ^{[+] \, \mu}_\partial (x)$ 
and in the Drell-Yan by $\Phi ^{[-] \, \mu}_\partial (x)$
\cite{Boer-2003}.
This fact leads to the sign change 
in the T-odd quark TMD functions:
\cite{collins-2002}
\begin{align}
f_{\text{SIDIS}} (x, k_T^{\, 2})_{\text{$T$-odd}} 
= - f_{\text{DY}} (x, k_T^{\, 2})_{\text{$T$-odd}} .
\label{eqn:tmd-sign}
\end{align}
The difference comes from the fact that the color interactions
are in the final state for the SIDIS and in the initial state
for the Drell-Yan as shown in Figs.\,\ref{fig:sidis-ex1} and 
\ref{fig:dy-ex1}, respectively.
It leads to the difference on the color-flow path 
between Eq.\,(\ref{eqn:wilson-2}) and Eq.\,(\ref{eqn:wilson-4}).

The color is confined in hadrons, so that the color degrees of freedom
usually does not appear explicitly in physical observables. 
However, depending on the color-flow direction, the T-odd TMDs are different
in sign. The TMD case is a rare and special occasion to investigate 
the color flow, namely the color degrees of freedom, in hadron physics.
It was predicated theoretically that the TMDs are different in sign 
between the SIDIS and the Drell-Yan process. 
In fact, there are already experimental indications on this new phenomenon 
in the Sivers functions.
About the experimental signatures on the sign change in the TMDs, 
it was suggested in the spin asymmetry of the reaction 
$\vec p + p \to W^\pm /Z^0+X$ 
by the STAR Collaboration \cite{star-sivers-2016}
and the spin asymmetry of $\pi^- + \vec p \to \mu^+ \mu^- +X$
by the COMPASS Collaboration \cite{compass-sivers-2017}.
Further confirmations on these effects are needed by future 
accurate experiments.

\section{Results on TMDs for tensor-polarized spin-1 hadrons}
\label{sec:3}

We derive possible quark TMDs for tensor-polarized spin-1 hadrons
in this section by 
the decomposition of the quark correlation function
in terms of kinematical factors in the Lorentz-invariant manner.
In particular, we find new terms associated with 
the lightcone vector $n$ in this work. 
First, we try to obtain all the possible
terms in the expansion of the quark TMD correlation function
in Sec.\,\ref{decomposition} by considering a tensor-polarized 
spin-1 hadron. 
Then, properties of each expansion term are discussed on
Hermiticity, parity, time reversal, chirality, and twist
in Sec.\,\ref{properties}.
Next, our guideline is explained for assigning various TMD notations 
in Sec.\,\ref{tmd-name-gauideline}, and
we show possible twist-2, 3, and 4 quark TMDs 
in Secs.\,\ref{twist-2}, \ref{twist-3}, and \ref{twist-4}, respectively.
A brief summary is given on the new TMDs and 
possible new fragmentation functions are explained
in Sec.\,\ref{comments-tmds-ffs}.
The new terms associated with $n$ modify the relations in the twist-2 TMDs,
which were obtained in the previous work \cite{bate2000}.
In addition, we show that there are 
new twist-3 and 4 TMDs in this work.

\subsection{Decomposition of quark correlation function}
\label{decomposition}

For spin-1/2 nucleon, the spin density matrix is parametrized 
with the spin vector which contains three parameters.
However, due to the spin-1 nature, the spin density matrix of 
the spin-1 hadron, such as the deuteron, is determined 
by spin tensor in addition to the spin vector.
There are five parameters in the spin tensor part, 
and the spin-vector part of spin-1 hadron is the same 
as the one of the spin-1/2 nucleon.

For expressing polarizations of the spin-1 hadron,
its density matrix is given by spin vector and tensor terms as
\cite{bate2000,Kumano:2020gfk}
\begin{align}
{\mathbold\rho} = \frac{1}{3} \left ( 1 + \frac{3}{2} \,  S_i {\mathbold\Sigma}_i
+ 3 \, T_{ij}  {\mathbold\Sigma}_{ij} \right ) .
\label{eqn:density-spin-1}
\end{align}
Here, ${\mathbold\Sigma}_i$ are $3 \times 3$ spin matrices 
for the spin-1 hadron,
and ${\mathbold\Sigma}_{ij}$ are spin tensors defined by
$
{\mathbold\Sigma}_{ij} = 
\left ( {\mathbold\Sigma}_i {\mathbold\Sigma}_j 
      + {\mathbold\Sigma}_j {\mathbold\Sigma}_i \right ) /2
- (2/3) \, {\mathbold I} \, \delta_{ij}  
$
with the $3 \times 3$ identity matrix ${\mathbold I}$.
The spin vector and tensor are parametrized as
\begin{align}
\! \! \! \! 
{\mathbold S} & = (S_{T}^x,\, S_{T}^y,\, S_L) ,
\nonumber \\
\! \! \! \! 
{\mathbold T}  & = \frac{1}{2} 
\left(
    \begin{array}{ccc}
     - \frac{2}{3} S_{LL} + S_{TT}^{xx}    & S_{TT}^{xy}  & S_{LT}^x  \\[+0.20cm]
     S_{TT}^{xy}  & - \frac{2}{3} S_{LL} - S_{TT}^{xx}    & S_{LT}^y  \\[+0.20cm]
     S_{LT}^x     &  S_{LT}^y              & \frac{4}{3} S_{LL}
    \end{array}
\right) ,
\label{eqn:spin-1-vector-tensor}
\end{align}
in the rest frame of the spin-1 hadron.
The parameters $S_{T}^x$ and $S_{T}^y$ indicate transverse polarizations
of the hadron, and $S_L$ does the longitudinal polarization.
The parameter $S_{LL}$ indicates the tensor polarization 
along the longitudinal axis as shown in Ref.\,\cite{bate2000}, 
and $S_{LT}^{x,y}$ ($S_{TT}^{xx,xy}$) indicate 
polarization differences along the axes 
between the longitudinal and transverse directions 
(along the transverse axes).
The linear polarizations are parts of the tensor polarizations.
These tensor and linear polarizations are schematically shown 
in the Appendix of Ref.\,\cite{bate2000}.

The covariant forms of $S^\mu$ and $T^{\mu\nu}$ of a spin-1 hadron
are generally expressed as \cite{bate2000,Kumano:2016ude}
\vspace{-0.20cm}
\begin{align}
S^\mu & = S_L \frac{P^+}{M} \bar n^\mu - S_L \frac{M}{2  P^+} n^\mu + S_T^\mu ,
\nonumber \\
T^{\mu\nu} & = \frac{1}{2} \left [ \frac{4}{3} S_{LL} \frac{(P^+)^2}{M^2} 
               \bar n^\mu \bar n^\nu 
          - \frac{2}{3} S_{LL} ( \bar n^{\{ \mu} n^{\nu \}} -g_T^{\mu\nu} )
\right.
\nonumber \\
& \! \hspace{-0.55cm} \! 
\left.
+ \frac{1}{3} S_{LL} \frac{M^2}{(P^+)^2}n^\mu n^\nu
+ \frac{P^+}{M} \bar n^{\{ \mu} S_{LT}^{\nu \}}
- \frac{M}{2 P^+} n^{\{ \mu} S_{LT}^{\nu \}}
+ S_{TT}^{\mu\nu} \right ],
\label{eqn:spin-1-tensor-1}
\\[-0.90cm] \nonumber
\end{align}
where $a^{\{ \mu} b^{\nu \}}$ indicates the symmetrized combination
$a^{\{ \mu} b^{\nu \}} = a^\mu b^\nu + a^\nu b^\mu$,
and $M$ is the hadron mass.

The general expression of the correlation function $\Phi (k, P, S, T)$ 
contains three parts: unpolarized, vector-polarized,
and tensor-polarized terms.
The unpolarized and vector-polarized distributions in the spin-1 hadron
are exactly the same as the relevant ones in the spin-1/2 nucleon; however,
we briefly explain past studies on the quark correlation function
for the nucleon.
First, the quark correlation function was decomposed
into 9 terms by imposing Hermiticity, parity invariance, and
time-reversal invariance in Ref.\,\cite{tangerman-prd-1995}.
Then, the quark TMD correlation function was decomposed in 
Refs.~\cite{Mulders:1995dh, tangerman-th} by introducing
T-odd terms, and there are 12 terms with coefficients denoted 
as $A_1- A_{12}$.
This decomposition was constructed with the vectors $P$, $S$ and $k$.

However, this decomposition was not complete
because the quark correlation function
depends on the vector $n$ through the gauge link $W(0, \xi | n)$. 
Therefore, the additional terms which depend on $n$ 
were investigated in Refs.~\cite{Goeke:2003az, Bacchetta:2004zf,
Goeke:2005hb, Metz:2008ib, Accardi:2009au}, and 20 new terms were
found and they are denoted as $B_1-B_{20}$. 
Therefore, there are 32 terms in total 
for the quark correlation function in the spin-1/2 nucleon.
These new terms of $n$ are important for understanding 
all the TMDs, collinear PDFs, and their relations.
Relations among the PDFs were derived by using 
the Lorentz invariant decomposition of the correlation function, 
so that they were often called ``Lorentz-invariance relations" 
\cite{Boer:1997nt}. These relations were modified 
due to the existence of these new terms
\cite{Kundu:2001pk, Goeke:2003az}.
Furthermore, another new twist-3 term appeared and it 
invalidated the Wandzura-Wilczek relation of the twist-2 level 
\cite{Metz:2008ib, Accardi:2009au}. 
On the other hand, these new terms also introduced new TMDs such as 
$e_{T}^{\perp}(x, k_T^{\, 2})$, $f_{T}^{\perp}(x, k_T^{\, 2})$  
and $g^{\perp}(x, k_T^{\, 2})$ for the nucleon
\cite{Goeke:2003az, Bacchetta:2004zf, Goeke:2005hb, Metz:2008ib}.
The unpolarized and vector polarized terms
in the quark correlation function of the spin-1 hadron are 
the same as the ones in the nucleon, and these 32 terms had
been already studied \cite{Goeke:2003az,Metz:2008ib}.

In this work, we focus on the tensor-polarized part
which does not exist in the spin-1/2 nucleon. 
The quark TMD correlation function of a spin-1 hadron 
was investigated in Ref.\,\cite{pd-drell-yan} by 
adding T-even terms to the 9 terms in the nucleon case
\cite{tangerman-prd-1995}.
The T-odd terms should be also considered together with
proper tensor polarizations \cite{bate2000}, 
so that there are 8 new terms in total in the tensor part,
where the relevant coefficients were named as $A_{13}$--$A_{20}$.
On the collinear PDFs of a spin-1 hadron, there are also studies
in possible hadron-tensor terms, helicity amplitudes, 
and operator forms \cite{Hoodbhoy:1988am,jlab-gluon-trans,jaffe-twist}.

However, the terms with the vector $n$, which are found
for the spin-1/2 nucleon, need to be added also in the formalism
of the spin-1 hadron, namely in the tensor-polarization part.
We formulate these new terms in this work 
to find possible TMDs. Including these $n$ terms, we express
the tensor part of quark correlation function $\Phi (k, P, T \, | n)$ 
for the spin-1 hadron as
\begin{widetext}
\vspace{-0.50cm}
\begin{align}
\! \! \! \!
\Phi(k, & P, T \, | n) = \frac{A_{13}}{M}  T_{kk} + \frac{A_{14}}{M^2} T_{kk} 
     \slashed{P}+ \frac{A_{15}}{M^2} T_{kk} \slashed{k}+  
     \frac{A_{16}}{M^3}  \sigma_{P k}   T_{kk} +A_{17}  T^{k \nu} 
     \gamma_\nu +\frac{A_{18}}{M} \sigma_{\nu P}  T^{k\nu}  
    +  \frac{A_{19}}{M} \sigma_{\nu k}  T^{k\nu}
\nonumber \\  
&  + \frac{A_{20}}{M^2} \varepsilon^{\mu\nu P k}  \gamma_{\mu} \gamma_5 T_{\nu k}
 + \frac{B_{21}M}{P\cdot n} T_{kn}  +\frac{B_{22}M^3}{(P\cdot n)^2}
T_{nn}+ \frac{B_{23}}{P\cdot n M} \varepsilon^{\mu kPn} T_{\mu k}(i\gamma_5)  
+\frac{B_{24}M}{(P\cdot n)^2} \varepsilon^{\mu kPn} T_{\mu n} (i\gamma_5)   
+\frac{B_{25}}{P\cdot n} \slashed{n} T_{kk} 
\nonumber \\ 
& + \frac{B_{26}M^2}{(P\cdot n)^2 } \slashed{n} T_{kn} 
 +\frac{B_{27}M^4}{(P\cdot n)^3 } \slashed{n} T_{nn}
+  \frac{B_{28}}{P\cdot n } \slashed{P} T_{kn}
+   \frac{B_{29}M^2}{(P\cdot n)^2 } \slashed{P} T_{nn}
+ \frac{B_{30}}{P\cdot n } \slashed{k} T_{kn}
+ \frac{B_{31}M^2}{(P\cdot n)^2 } \slashed{k} T_{nn} 
+ \frac{B_{32}M^2}{P\cdot n } \gamma_\mu  T^{\mu n} 
\nonumber \\
& + \frac{B_{33} }{P \cdot n} \varepsilon^{\mu \nu P k}
                      \gamma_{\mu} \gamma_5  T_{\nu n}  
+ \frac{B_{34}}{P \cdot n} \varepsilon^{\mu \nu P n} 
                      \gamma_{\mu} \gamma_5  T_{\nu k}  
+ \frac{B_{35} M^2}{(P \cdot n)^2} \varepsilon^{\mu \nu P n} 
                      \gamma_{\mu} \gamma_5  T_{\nu n}   
+   \frac{B_{36} }{P \cdot n M^2} \varepsilon^{\mu k P n}
                      \gamma_{\mu} \gamma_5  T_{kk} 
\nonumber \\
& +\frac{B_{37}}{(P \cdot n)^2}  \varepsilon^{\mu k P n}
                      \gamma_{\mu} \gamma_5  T_{kn} 
 +\frac{B_{38} M^2 }{(P \cdot n)^3 } \varepsilon^{\mu k P n}
                      \gamma_{\mu} \gamma_5  T_{nn}
 +  \frac{B_{39} }{(P \cdot n)^2}  \slashed{n} 
                      \gamma_5 T_{\mu k} \varepsilon^{\mu k P n}  
+ \frac{B_{40} M^2}{(P \cdot n)^3}  \slashed{n} 
                      \gamma_5 T_{\mu n} \varepsilon^{\mu k P n}
\nonumber \\ 
& + \frac{B_{41}}{P\cdot nM}   \sigma_{P k} T_{kn} 
+ \frac{B_{42}M}{(P\cdot n)^2} \sigma_{Pk} T_{nn}
+ \frac{B_{43}}{P\cdot n M}   \sigma_{P n}  T_{kk}  
+\frac{B_{44}M}{(P\cdot n )^2} \sigma_{P n} T_{kn} 
+ \frac{B_{45}M^3}{(P\cdot n)^3 } \sigma_{Pn}  T_{nn}
+ \frac{B_{46}}{P\cdot n M}   \sigma_{k n} T_{kk}
\nonumber \\
& + \frac{B_{47}M}{(P\cdot n)^2 } \sigma_{k n} T_{kn}
+ \frac{B_{48}M^3}{(P\cdot n)^3 } \sigma_{kn} T_{nn}
+\frac{B_{49}M}{P\cdot n } \sigma_{\mu n} T^{\mu k} 
+ \frac{B_{50}M^3}{(P\cdot n)^2 } \sigma_{\mu n} T^{\mu n} 
+ \frac{B_{51}M}{P\cdot n } \sigma_{\mu P} T^{\mu n}
+ \frac{B_{52}M}{P\cdot n } \sigma_{\mu k}  T^{\mu n} ,
\label{eqn:cork4}
\end{align} 
\end{widetext}
where the notation $X_{\mu k} \equiv X_{\mu \nu} k^{\nu}$ 
is used for brevity with the tensor $X$ as 
$\sigma^{\mu \nu}= i \left[ \gamma^{\mu} , \gamma^{\nu} \right]/2$, 
$T^{\mu\nu}$, or the antisymmetric tensor 
$\varepsilon^{\mu \nu  \alpha \beta}$, 
and $k$ could be replaced by $n$ or $P$. 
We listed only the tensor terms proportional to
the tensor polarization $T^{\mu\nu}$ in Eq.\,(\ref{eqn:spin-1-tensor-1}).
Here, we use the convention 
$\varepsilon^{0123}=+1$ so as to agree with expressions
in Ref.\,\cite{bate2000}. 
In deriving this equation, the Hermiticity and  parity-invariance
relations in Table \ref{table:H-P-T} are imposed
for the correlation function;
however, the time-reversal invariance is not 
a necessary condition due to the existence of the gauge link.

The first 8 terms ($A_{13}$--$A_{20}$) were already obtained
in Ref.\,\cite{bate2000}, and they generated 
all the leading-twist TMDs.
There are 40 terms in the tensor part of the quark correlation function, 
and 32 of them ($B_{21}$--$B_{52}$) are dependent on the vector $n$.
Therefore, the new terms, which we found in this work, 
are these 32 terms $B_{21}$--$B_{52}$.

In general, the coefficients $A_i$ ($i=1$--$20$) and $B_i$ ($i=1$--$52$) 
depend on the scalars $k \cdot P$, $k^2$, $P \cdot n$ and $k \cdot n$. 
In order to keep $\Phi (k, P, T \, | n)$ invariant when the vector $n$
is replaced by $\lambda n$ as a scale change, 
$A_i$ and $B_i$ should be functions 
of $k^2$ and the ratios, $k \cdot n/  P \cdot n$ and $k \cdot P$
\cite{ Accardi:2009au}.
The quark and hadron momenta $k$ and $P$ are expressed 
by two lightlike vectors $n$ and $\bar{n}$ as
\begin{align}
P^\mu & = P^+ \bar n^\mu + \frac{M^2}{2  P^+} n^\mu , 
\nonumber \\
k^\mu & = x P^+ \bar n^\mu + \frac{M^2(\sigma-x)}{2  P^+} n^\mu + k_T^\mu,
\label{eqn:kp-momenta}
\end{align}
where $k_T^{\, 2} (= - \vec k_T^{\, 2})$, $\sigma$, and $\tau$ are given by
\begin{align}
k_T^{\, 2}  =  (\tau+x^2-x\sigma) M^2, \ \ 
\sigma \equiv \frac{2k \cdot P}{ M^2}, \ \ 
\tau \equiv \frac{k^2}{M^2}.
\label{eqn:mo2}
\end{align}
Here, $x$ is the lightcone momentum fraction carried by the quark.

The $k_T$-dependent correlation function is obtained 
by integrating $\Phi (k, P,  T | n)$ over $k^{-}$,
\begin{align}
\Phi(x, k_T, T ) =\int dk^{-}  \Phi(P, k, T \, | n) .
\label{eqn:mo1}
\end{align}
The TMD correlation function 
$\Phi(x, k_T, T )$ 
is used to describe the hard processes such as 
the semi-inclusive DIS and Drell-Yan process.
Using the TMD correlation function of Eq.\,(\ref{eqn:mo1}), 
we define the trace of TMD function by
\begin{align}
\Phi^{\left[ \Gamma \right]} (x, k_T, T) \equiv 
\frac{1}{2} \, \text{Tr} \left[ \, 
\Phi(x, k_T, T ) 
\Gamma \, \right] ,
\label{eqn:trace-tmds}
\end{align} 
where $\Gamma$ is a gamma matrix. 
We reiterate that this correction function is only for 
the tensor-polarization ($T$) part, and the unpolarized and 
vector-polarized ($S$) terms are not included 
because they have been already investigated in previous works
\cite{Goeke:2005hb,Metz:2008ib}.

\subsection{Properties of Hermiticity, parity, time reversal, chirality,
and twist}
\label{properties}

Each term of the expansion in Eq.\,(\ref{eqn:cork4}) satisfies
the Hermiticity and parity invariance in Table \ref{table:H-P-T}.
The time-reversal invariance is not imposed because of
the active role of the gauge link in the TMDs.
We explain the details on the conditions of
Hermiticity, parity invariance, time-reversal invariance, 
chirality, and twist in the following.

\vspace{0.10cm}
\noindent
{\bf [Hermiticity]}\\
\noindent
The Hermiticity condition
$\Phi ^ \dagger (V, A, T) 
= \gamma^0 \, \Phi (V, A, T) \gamma^0$,
where $V$ is a Lorentz vector, $A$ is an axial vector,
and $T$ is a tensor, is satisfied because of
the relations $(\Gamma)^\dagger = \gamma^0 \Gamma \gamma^0$ 
by taking $\Gamma$ as
\begin{align}
{\bf 1},\ \gamma^\mu,\
    \gamma^\mu\gamma_5,\  i \gamma_5,\ \sigma^{\mu\nu} ,
\label{eqn:Hermiticity-list}
\end{align} 
where ${\bf 1}$ is the $4 \times 4$ identity matrix.

\vspace{0.10cm}
\noindent
{\bf [Parity invariance]}\\
\noindent
The parity-invariance relation indicates
$\Phi (V,A,T) 
= \gamma^0 \, \Phi (\bar V, - \bar A, \bar T) \gamma^0$,
which is satisfied, for example, 
because of the relation
$\gamma^0 \slashed{\bar V} \gamma^0 = \slashed{V}$ 
for the vector $V^\mu$ and 
$\gamma^0 (- \gamma_5 \slashed{\bar A}) \gamma^0 = \gamma_5 \slashed{A}$
for the axial vector $A^\mu$.
We may note that the term 
$\varepsilon^{\mu XYZ} = \varepsilon^{\mu\nu\alpha\beta} X_\nu Y_\alpha Z_\beta$
is an axial vector, so that $\gamma_5 \gamma_\mu \varepsilon^{\mu XYZ}$
and $i \gamma_5 \varepsilon^{VXYZ}$ terms satisfy the parity invariance.
Here, $X$, $Y$, and $Z$ are Lorentz vectors. 
In fact, we have the relation
$\gamma^0 (i \gamma_5 \varepsilon^{\bar V \bar X \bar Y \bar Z}) \gamma^0 
 = i \gamma_5 \varepsilon^{VXYZ}$.
However, the pseudoscalar term $i \gamma_5 $ is not allowed due 
to the relation $\gamma^0 (i \gamma_5) \gamma^0 = - i \gamma_5$.
In the same way, the pseudoscalar term $\varepsilon^{VXYZ}$ is not allowed.
The term $\varepsilon^{AXYZ}$ with the axial vector $A^\mu$
exists because of 
$\gamma^0 (\varepsilon^{(-\bar A) \bar X \bar Y \bar Z}) \gamma^0 
  = \varepsilon^{A XYZ}$.
The term $\sigma^{XY}= \sigma^{\mu\nu} X_{\mu} Y_{\nu}$
is allowed under the parity invariance because of 
$\gamma^0 \sigma^{\bar X \bar Y} \gamma^0 = \sigma^{XY}$, 
so that various $\sigma^{\mu\nu}$ terms exist in Eq.\,(\ref{eqn:cork4}).
These discussions are summarized as the properties under the parity 
transformation: 
\begin{alignat}{2}
& \text{P-even:} & \ \, & {\bf 1},\  \slashed{V},\ 
                       \gamma_5 \slashed{A},\ 
                       i \gamma_5 V \cdot A,\ 
                       i \gamma_5 \varepsilon^{VXYZ},\ 
                       \gamma_5 \gamma_\mu \varepsilon^{\mu XYZ},\                   
\nonumber \\
&  \  & \  &  
           \gamma_\mu \varepsilon^{\mu XYZ},\ 
              \varepsilon^{AXYZ},\ \sigma^{XY},\ 
              i\gamma_5 \sigma^{AX},\ 
              \cdots,
\nonumber \\
& \text{P-odd:} & \ \, &  i \gamma_5,\ 
         \slashed{A},\ \varepsilon^{VXYZ},\ 
      \gamma_\mu \varepsilon^{\mu XYZ},\ \sigma^{AX},\ 
      i \gamma_5 \sigma^{XY},\ 
\cdots .
\label{eqn:p-even-odd-list}
\end{alignat} 

\vspace{-0.30cm}
\noindent
Since the parity invariance is imposed in the correlation function,
the parity-odd terms do not appear in Eq.\,(\ref{eqn:cork4}).

\vspace{0.10cm}
\noindent
{\bf [Time reversal]}\\
\noindent
The time-reversal property is given in Table \ref{table:H-P-T} as 
$\Phi ^* (V,A,T) 
= {\pazocal T} \, \Phi (\bar V,\bar A, \bar T) \, {\pazocal T}^{-1}$ 
where ${\pazocal T}= -i\gamma_5 C = i \gamma^1 \gamma^3 
       = {\pazocal T}^\dagger = {\pazocal T}^{-1} = - {\pazocal T}^*$
\cite{bd-book-II}.
Because of the $\gamma$-matrix relation 
${\pazocal T} \gamma^\mu {\pazocal T}^{-1} 
  =\gamma^{\mu T} =\bar\gamma^{\mu\, *}$, 
the term $\slashed{V}=V_\mu \gamma^\mu$ satisfies the time-reversal relation
${\pazocal T} \slashed{\bar V} {\pazocal T}^{-1}
  = \slashed{V}^*$,
so that it is called T-even term. 
In the same way, the scalar term ({\bf 1} without a $\gamma$ matrix)
and the other ones
($\gamma_5 \slashed{A}$,\ $\gamma_5 \slashed{V}$,\ 
 $ i \gamma_5 \varepsilon^{VXYZ}$, $ i \gamma_5 \sigma^{AX}$)
satisfy the time-reversal-invariance relation,
and they are T-even terms.
We may note that the imaginary $i$ exists as $i \gamma_5$,
whereas it does not exist in $\gamma_5 \gamma^\mu$, because
of the Hermiticity requirement $\Phi^\dagger = \gamma^0 \Phi \gamma^0$.

However, the time-reversal relation is not satisfied for the terms
with $\sigma^{XY}$, $\varepsilon^{VXYZ}$, and the others.
For example, since the tensor $\sigma^{\mu\nu}$  has the property
${\pazocal T} \sigma^{\mu\nu} {\pazocal T}^{-1} 
  = - (\bar \sigma^{\mu\nu})^*$
under the time reversal, the term $\sigma^{XY}$
has the relation
${\pazocal T} \sigma^{\bar X \bar Y} {\pazocal T}^{-1} 
= - (\sigma^{X Y})^* $
with the negative sign.
This relation is same for the $i \gamma_5 \sigma^{XY}$ term.
Therefore, they are called T-odd terms due to the negative sign.
They are summarized as follows:
\begin{alignat}{2}
& \! \!
\text{T-even:} & \ \, & {\bf 1},\  \slashed{V},\ 
                     \gamma_5 \slashed{A},\ 
                    i \gamma_5 \varepsilon^{VXYZ},\ 
                    i \gamma_5 \sigma^{AX},\  \cdots ,
\nonumber \\
& \! \! 
\text{T-odd:} & \ \, & i \gamma_5,\ \sigma^{XY},\ 
              i \gamma_5 \sigma^{XY},\ 
              i \gamma_5 V \cdot A ,\ 
                \varepsilon^{VXYZ},\ 
\nonumber \\
&  \  & \  &
              \gamma_\mu \varepsilon^{\mu XYZ} ,\ 
              \gamma_5 \gamma_\mu \varepsilon^{\mu XYZ},\ \cdots .
\label{eqn:t-even-odd-list}
\end{alignat} 
\noindent
Among them, the terms $i \gamma_5$, $i \gamma_5 \sigma^{XY}$,
$\varepsilon^{VXYZ}$, and $\gamma_\mu \varepsilon^{\mu XYZ}$
are  ruled out by the parity invariance, so that
they do not appear in Eq.\,(\ref{eqn:cork4}).
From this time-reversal classification,
the expansion terms of Eq.\,(\ref{eqn:cork4})
have the T-even and T-odd properties as
\begin{alignat}{2}
& \text{T-even terms:} & \ \, & A_{13-15}, \  A_{17}, \ B_{21-32},
\nonumber \\
& \text{T-odd terms:}  & \ \, & A_{16},\ A_{18-20},\ B_{33-52}.
\label{eqn:t-even-odd-terms}
\end{alignat} 
Just in case, we also list the time-reversal properties
in the unpolarized and vector polarization cases 
in Ref.\,\cite{Goeke:2005hb} as
\begin{alignat}{2}
& \text{T-even terms:} & \ \, & A_{1-3}, \  A_{6-11}, \ B_{1}, \ B_{11-20},
\nonumber \\
& \text{T-odd terms:}  & \ \, & A_{4-5},\ A_{12},\ B_{2-10}.
\label{eqn:t-even-odd-terms-2}
\end{alignat} 

\vspace{0.10cm}
\noindent
{\bf [Chirality]}\\
\noindent
The TMDs and PDFs are also classified by the chiral property.
Structure functions of a hadron are given by the imaginary
part of forward scattering amplitudes by the optical theorem,
so that the TMDs and PDFs are expressed by parton-hadron forward 
scattering amplitudes in Fig.\,\ref{fig:correlation-q}.
The quark transversity distribution $h_1$ (or denoted as 
$\Delta_T q$)
is associated with the quark spin-flip ($\lambda_i=+$, $\lambda_f=-$) amplitude, 
so that it is called a chiral-odd distribution.
This distribution is defined by the matrix element 
with the $\gamma$ matrix term, $i\gamma_5 \sigma^{\mu\nu}$,
as shown in Eq.\,(13) of Ref.\,\cite{Kumano:2020gfk}.
At high energies, the helicity is conserved 
for the vector ($\gamma^\mu$) and axial-vector ($\gamma_5 \gamma^\mu$)
interactions.
We define the right-handed and left-handed fermion states
as $\psi_R = 1/2 (1-\gamma_5) \psi$ and $\psi_L = 1/2 (1+\gamma_5) \psi$,
which correspond to the helicity $+1$ and $-1$ states, respectively,
at high energies where the fermion masses are neglected.
For example, the relation 
$\bar \psi \gamma^\mu \psi = \bar\psi_L \gamma^\mu \psi_L 
  + \bar\psi_R \gamma^\mu \psi_R$ 
is satisfied due to the anticommutation relation
$\{ \gamma_5, \gamma^\mu \} =0$
and there is no cross term like $\bar\psi_L \gamma^\mu \psi_R$.
This relation is also the same for the axial vector current $\gamma_5 \gamma^\mu$.
These facts suggest that the quark helicities should be conserved in high-energy
strong, electromagnetic, and weak interactions.
However, the situation is different in terms with even number 
of $\gamma$ matrices. The helicity is not conserved for
scalar ({\boldmath$1$}), axial ($\gamma_5$), tensor $\sigma^{\mu\nu}$, 
and axial-tensor ($i \gamma_5\sigma^{\mu\nu}$) terms. 
For example, the relation becomes
$\bar \psi {\bf 1} \psi = \bar\psi_L  \psi_R + \bar\psi_R  \psi_L$.
Therefore, the chiral-even and chiral-odd $\gamma$ matrices are 
classified as
\begin{alignat}{2}
& \text{$\chi$-even:} & \ \, & \gamma^\mu, \  \gamma_5 \gamma^\mu, 
\nonumber \\
& \text{$\chi$-odd:}  & \ \, & {\bf 1}, \  i \gamma_5, \ 
                      \sigma^{\mu\nu},\ i \gamma_5\sigma^{\mu\nu} .
\label{eqn:chi-even-odd-list}
\end{alignat} 
\noindent
Using this classification on the chiral property, we obtain
the chiral-even and chiral-odd terms of Eq.\,(\ref{eqn:cork4}) as
\begin{alignat}{2}
& \! \! 
\text{$\chi$-even terms:} & \ \, & A_{14-15},\ A_{17},\ A_{20},\ B_{25-40},
\nonumber \\
& \! \! 
\text{$\chi$-odd terms:}  & \ \, & A_{13},\ A_{16},\ A_{18-19},\ 
                                    A_{21-24},\ B_{41-52}.
\label{eqn:chi-even-odd-terms}
\end{alignat} 
The chiral properties in the unpolarized and vector polarization 
cases in Ref.\,\cite{Goeke:2005hb} are also listed as
\begin{alignat}{2}
& \! \! \!  
\text{$\chi$-even terms:} & \ & A_{2-3},\ A_{6-8},\ A_{12},\ 
                                   B_1,\ B_{4},\ B_{7-14},
\nonumber \\
& \! \! \!
\text{$\chi$-odd terms:}  & \ & A_{1},\ A_{4-5},\ A_{9-11},\
                                   B_{2-3},\ B_{5-6},\ B_{15-20}.
\label{eqn:chi-even-odd-terms-2}
\end{alignat} 

\vspace{0.10cm}
\noindent
[{\bf Twist of the TMDs}]\\
\noindent
Let us take the frame where 
the hadron's longitudinal momentum is much larger than the hadron mass, 
namely $P^+ \gg M$, by taking the hadron momentum direction as the third axis
as given in Eq.\,(\ref{eqn:kp-momenta}),
and then consider the charged-lepton deep inelastic scattering
from the hadron. This frame could correspond
to the center-of-momentum
frame between the virtual photon emitted from the lepton and the hadron.
Then, $P^+$ is related to the scale $Q^2$ by the relation
$P^+ \simeq \sqrt{Q^2/(2x(1+x))} \sim O (Q)$
by neglecting the hadron mass.

In the operator-product expansion, the structure functions are
classified by the twist, which is the operator mass dimension
minus the operator spin \cite{jaffe-twist}.
This twist controls the scaling behavior of the structure functions
as $Q^2$ becomes larger.
The leading-twist is two and the leading-twist structure functions
or the TMDs in this work have scaling behavior with the order of 
$O(1)$, and the twist-3 and 4 ones are $O(1/Q)$ and $O(1/Q^2)$,
respectively. Because of $P^+ \sim O(Q)$, the leading twist-2 TMDs
are defined in the TMD correlation functions as the terms of $O(1)$
as shown in Sec.\,\ref{twist-2}.
On the other hand, the twist-3 and 4 TMDs are given as the terms
of $O(1/P^+)$ and $O(1/(P^+)^2)$
as shown in Sec.\,\ref{twist-3} and Sec.\,\ref{twist-4}.

\subsection{Guideline for assigning TMD notations}
\label{tmd-name-gauideline}

We follow the TMD notations of Refs.\,\cite{Goeke:2005hb,bate2000}
as much as possible; however, there are new TMDs which need to be defined
in this work. The twist-2 TMDs were already named for the tensor-polarized
spin-1 hadron in Ref.\,\cite{bate2000}, and the same notations are 
used in twist 2.
However, all the twist-3 and twist-4 TMDs are new ones
for the the tensor-polarized spin-1 hadron, so that new names 
should be assigned.
In the twist-3 part, our notations are given in the similar spirit to 
to the twist-3 TMDs of the spin-1/2 nucleon in Ref.\,\cite{Goeke:2005hb}.
In twist 4, the TMD names are given by replacing all the twist-2 
subscripts $1$ (such as in $f_{1LL}$) by twist-4 ones $3$ ($f_{3LL}$).
The general guideline is the following.
\begin{enumerate}
\setlength{\leftskip}{+0.40cm}
\setlength{\itemsep}{-0.05cm} 
\item
The TMD function names $f$,  $g$, and $h$ are assigned
to the unpolarized, longitudinal, and transverse quark 
polarizations by taking traces of Eq.\,(\ref{eqn:trace-tmds}) with 
$\gamma^+$, $\gamma^+ \gamma_5$, and $i \sigma^{i+}\gamma_5$ 
(or $\sigma^{i+}$), respectively, 
in the twist-2 case.
The quark operators $\bar\psi \gamma^+ \psi$, 
$\bar\psi \gamma^+ \gamma_5 \psi$, and
$\bar\psi i \sigma^{i+} \gamma_5 \psi$
are related to the unpolarized (U), longitudinally polarized,
and transvsere polarization (T) of a quark in the twist-2 case
as given in Ref.\,\cite{Kumano:2020gfk}.
However, the twist-3 and twist-4 TMDs indicate three and four parton
correlations, so that they are not related to 
these quark distributions. 
Therefore, the distributions $f$, $g$, and $h$ are assigned depending
on the operator forms as shown in Tables 
\ref{table:twist-3-tmd-list} and \ref{table:twist-4-tmd-list}.
\item 
The subscript 1 in the TMDs, such as $f_{1LL}$, is assigned
for the twist-2 TMDs. The subscript 3 is used for
the twist-4 TMDs like $f_{3LL}$; 
however, the subscript 2 is not conventionally 
used for expressing the twist-3 TMDs.
\item 
The subscripts $LL$, $SL$, and $TT$ are given if TMDs 
appear with the spin parameters $S_{LL}$, $S_{LT}$, and $S_{TT}$,
respectively, in the traces of the TMD correlation functions
of Eq.\,(\ref{eqn:trace-tmds}). For example, 
$f_{1LL}$, $f_{1LT}$, and $f_{1LT}$ are defined in 
Eq.\,(\ref{eqn:cork-2}) in this way.
\item
The superscript $\perp$ ($F^\perp$)
is given if a TMD exists with the partonic transverse momentum 
$k_T^i$ ($i=1,2$). In addition, the superscript $\prime$ ($F^\prime$)
is assigned if two similar polarization and momentum factors
exist within the same TMD correlation function $\Phi^{[\Gamma]}$.
If both $F^\perp$ and $F^\prime$ exist, the superscript $\perp$ 
is assigned for the term with the partonic transverse-momentum ($k_T$)
term of the order of $(k_T)^2$ or $(k_T)^3$ in traces 
of TMD correlation functions.
An example is $h_{1 LT}^\perp$ in Eq.\,(\ref{eqn:cork-2}).
However, although the corresponding $F^\prime$ does not exist, the $F^\perp$
could be used even in the order of $(k_T)^1$, 
and its example is $h_{1 LL}^\perp$.
The TMDs with $\prime$ are assigned 
in the leading order of $k_T$, namely $O((k_T)^0)$ or  $O((k_T)^1)$.
An example is $h_{1LT}^{\prime}$ in Eq.\,(\ref{eqn:cork-2}).
In general, new TMDs $F$ are defined from the TMDs
$F^\perp$ and $F^{\,\prime}$ by the relation of 
Eq.\,(\ref{eqn:tmd-prime-f}), so that the TMD lists are shown
by the two-independent TMDs $F$ and $F^\perp$ without $F^{\,\prime}$
in Tables \ref{table:twist-2-tmd-list}, \ref{table:twist-3-tmd-list}, and 
\ref{table:twist-4-tmd-list}.
If both $F$ and $F^\perp$ exist, the meaning of $F$ and $F^\perp$ 
is more clearly shown in Eq.\,(\ref{eqn:redefine-tmd}).
The function $F^\perp$ [$e.g.$ $h_{1 LT}^\perp$ in Eq.\,(\ref{eqn:redefine-tmd})]
is given as the term with a kinematical factor which vanishes 
by the $\vec k_T$ integration. 
The other function $F$ ($e.g.$ $h_{1 LT}$)
is assigned for the remaining part.
\item
There are exceptions for the above assignment of $\perp$.
The superscript $\perp$ is not written conventionally for 
$f_{1LT}$, $f_{1TT}$, $g_{1LT}$, $g_{1TT}$, 
(also the twist-4 distributions 
$f_{3LT}$, $f_{3TT}$, $g_{3LT}$, and $g_{3TT}$), 
although they accompany $k_T$ factors 
in the correlation functions.
\item
The superscript $\prime$ is not written
if similar functions exist in separate correlation functions 
$\Phi^{[\Gamma_1]}$ and $\Phi^{[\Gamma_2]}$.
In the traces for the twist-3 TMD correlation functions 
$\Phi^{[\sigma^{-+}]}$ and $\Phi^{[\sigma^{ij}]}$
in Eq.\,(\ref{eqn:cork-3-3}), 
the $k_T$ dependence is the same order $O((k_T)^1)$ for $h_{LT}^\perp$,
so one may assign $h_{LT}^{\perp (1)}$ and $h_{LT}^{\perp (2)}$.
Similar expressions appeared in the twist-3 part of the nucleon,
and they were already named as $h_T$ and $h_T^\perp$ 
in Ref.\,\cite{Goeke:2005hb}.
Following such a convention, we write them as 
$h_{LT}$ and $h_{LT}^{\perp}$ in Eq.\,(\ref{eqn:cork-3-3}).
In this equation, $h_{TT}$ and $h_{TT}^{\perp}$ are also written
in the same manner in Eq.\,(\ref{eqn:cork-3-3})
although they have the same dependence of $O((k_T)^0)$.
In the same way, $e_{LT}$, $e_{LT}^\perp$, $e_{TT}$, and $e_{TT}^\perp$
are assigned in $\Phi^{[1]}$ and $\Phi^{[i\gamma_5]}$.
\vspace{-0.20cm}
\end{enumerate}

\subsection{Twist-2 TMDs for a tensor-polarized spin-1 hadron}
\label{twist-2}

The leading twist TMDs for a tensor-polarized spin-1 hadron
are defined by taking
$\Gamma = \gamma^+$, $\gamma^+ \gamma_5$, and $\sigma^{i+}$
in Eq.\,(\ref{eqn:trace-tmds}), and we obtain
\vspace{-0.20cm}
\begin{align}
\displaybreak[2]
\Phi^{[\gamma^+]} (x, k_T, T) & = f_{1LL}(x, k^{\,2}_T) S_{LL}
   - f_{1LT}(x, k^{\,2}_T) \frac{S_{LT}\cdot k_T}{M} 
\nonumber \\\displaybreak[2]
& \ \hspace{-1.7cm} \ 
+f_{1TT}(x, k^{\,2}_T) \frac{k_T\cdot S_{TT}\cdot k_T}{M^2} ,
\nonumber \\
\Phi^{[\gamma^+ \gamma_5]} (x, k_T, T) &  =  
g_{1LT}(x, k^{\,2}_T) 
     \frac{S_{LT\, \mu} \, 
     \varepsilon_{T}^{\mu\nu} \, k_{T\, \nu}}{M} 
\nonumber \\\displaybreak[2]
& \ \hspace{-1.7cm} \
+ g_{1TT}(x, k^{\,2}_T) \frac{S_{TT\, \mu \rho} \, k_{T}^{\rho} \,
     \varepsilon_{T}^{\mu\nu} k_{T\, \nu}}{M^2} ,
\nonumber \\\displaybreak[2]
\Phi^{[\sigma^{i+}]} (x, k_T, T) & =  
h^{\perp}_{1LL}(x, k^{\,2}_T) \frac{S_{LL} k_T^i}{M}
+ h^{\prime}_{1LT} (x, k^{\,2}_T) S_{LT}^i 
\nonumber \\\displaybreak[2]
& \ \hspace{-1.7cm} \
- h_{1LT}^{\perp}(x, k^{\,2}_T) \frac{ k_{T}^i  S_{LT}\cdot k_{T}}{M^2} 
- h_{1TT}^{\prime} (x, k^{\,2}_T) \frac{S_{TT}^{ i j} k_{T j} }{M}
\nonumber \\\displaybreak[2]
& \ \hspace{-1.7cm} \
+ h_{1TT}^{\perp}(x, k^{\,2}_T) \frac{k_T\cdot S_{TT}\cdot k_T}{M^2} 
    \frac{k_T^i}{M},
\label{eqn:cork-2}
\end{align} 
where $i$ and $j$ indicate the transverse indices $i=1,\, 2$ ($j=1,\, 2$),
$\varepsilon_{T}^{\mu\nu} 
  =\varepsilon^{\mu \nu \alpha \beta} \bar{n}_{\alpha} n_{\beta}$
is used with the convention $\varepsilon^{0123}=1$, and 
$S_{LT} \cdot k_T$ and $k_T\cdot S_{TT}\cdot k_T$ are defined by
$S_{LT} \cdot k_T = S_{LT}^i k_{Ti} = - S_{LT}^i k_{T}^i$ and 
$k_T\cdot S_{TT}\cdot k_T = k_{Ti} \, S_{TT}^{ij} \, k_{Tj}$.
Here, we follow the notations of Ref.\,\cite{bate2000} 
for the TMD expressions in twist 2.
In Ref.\,\cite{bate2000}, the trace with $i \sigma^{i+} \gamma_5$
was taken instead of $\sigma^{i+}$; however, both formalisms
are equivalent by considering the relation
$i \sigma^{\mu\nu} \gamma_5
= - \varepsilon^{\mu\nu\alpha\beta} \sigma_{\alpha\beta}/2$
\cite{Itzeykson-Zuber}.
Therefore, if $\Phi^{[i \sigma^{i+} \gamma_5]}$ is calculated,
the same equation is obtained by the replacements
$X^i \to \varepsilon_T^{ij} X_j$ with $X^i =k_T^i$, $S_{LT}^i$, 
and $S_{TT}^{ij} k_{Tj}$ in $\Phi^{[\sigma^{i+}]}$
of Eq.\,\,(\ref{eqn:cork-2}).
There are 10 TMDs in the leading-twist level, 
as already found in Ref.\,\cite{bate2000}. 
However, their relations to the expansion coefficients 
are modified due to the existence of the new terms $B_{21-52}$
associated with the tensor structure and the lightlike vector $n$,
as we find in Eqs.\,(\ref{eqn:twist-2-f}), (\ref{eqn:twist-2-g}),
and (\ref{eqn:twist-2-h}).

The two TMDs $h^{\prime}_{1LT}$ ($h^{\prime}_{1TT}$) and
$h_{1LT}^{\perp}$ ($h_{1TT}^{\perp}$) are similar notations.
Because of the relation
\begin{align}
k_T^i k_T \cdot S_{LT} 
= 
\varepsilon_{T}^{ij} k_{T j}  k_{T l} \varepsilon_{T}^{lm} S_{LT m}
+ k_T^2 S_{LT}^i ,
\label{eqn:cork-mat}
\end{align}
the other functions $h_{1LT}$ and $h_{1TT}$ could be defined
instead of $h^\prime_{1LT}$ and $h^\prime_{1TT}$.
In fact, the correlation function $\Phi^{[\sigma^{i+}]}$
in Eq.\,(\ref{eqn:cork-2}) is rewritten as
\begin{align}
& \! \! \! \!
\Phi^{[\sigma^{i+}]} 
= h^\perp_{1LL} \frac{S_{LL} k_T^i}{M}
\nonumber \\\displaybreak[2]
& \! \! \! \! \! \! \! 
+ h_{1LT} S_{LT}^i 
+ h_{1LT}^\perp \frac{S_{LT}^j k_T^i k_{T}^j - S_{LT}^i \vec k_T^2 /2}{M^2} 
\nonumber \\\displaybreak[2]
& \! \! \! \! \! \! \!
+ h_{1TT} \frac{S_{TT}^{ i j} k_{T}^j }{M}
+ h_{1TT}^\perp \frac{(  S_{TT}^{lj} k_T^i k_{T}^l 
         - S_{TT}^{ij} \vec k_T^2 /2 ) k_{T}^j}{M^3} ,
\label{eqn:redefine-tmd}
\end{align}
Here, we define the new functions without $\prime$ and $\perp$ as
\begin{align}
F (x, k_T^{\, 2}) \equiv F^{\,\prime} (x, k_T^{\, 2})
 - \frac{k_T^{\, 2}} {2M^2} \, F^{\perp} (x, k^{\, 2}_T) .
\label{eqn:tmd-prime-f}
\end{align}
where $F = h_{1LT}$ and $h_{1TT}$ in the twist 2
and $k_T^{\, 2}= - \vec k_T^{\, 2}$,
as this relation was written for the unpolarized TMD 
($f$, $f^\prime$, and $f^\perp$)
in Ref.\,\cite{br-book}. 
We note in Eq.\,(\ref{eqn:tmd-prime-f}) that 
the $h_{1LT}^\perp$ and $h_{1TT}^\perp$ terms 
vanish by the $\vec k_T$ integration.
It leads to the sum rule for $f_{1LT}$ in Eq.\,(\ref{eqn:cork-5}).
Therefore, two of these three functions 
$h_{1LT}$, $h^\prime_{1LT}$, and $h_{1LT}^{\perp}$
(also $h_{1TT}$, $h^\prime_{1TT}$, and $h_{1TT}^{\perp}$)
are independent, so that one could choose two of them depending on
one's preference in defining the TMDs in Eq.\,(\ref{eqn:cork-2}).
Similar relations appear in twist-3 and twist-4 cases,
so that we use Eq.\,(\ref{eqn:tmd-prime-f}) 
as the general relation for the TMD $F$ 
in terms of $F^{\,\prime}$ and $F^\perp$.

Calculating traces in Eq.\,(\ref{eqn:trace-tmds})
with the new correlation function of Eq.\,(\ref{eqn:cork4}),
we express the twist-2 TMDs 
in Eq.\,(\ref{eqn:cork-2}) in terms of the coefficients $A_i$ and $B_i$.
First, the unpolarized quark TMDs in $\Phi^{[\gamma^+]}$ are given as
\begin{align}
f_{1LL}(x, k_T^{\, 2}) & 
  =  \frac{P^+}{3} \int dk^- \left[ (A_{14}+xA_{15}) \tau_x 
\right.
\nonumber \\
& \! \hspace{-1.6cm} \! 
\left.
             + 2( A_{17} + B_{28}+xB_{30} )(\sigma -2x)  
             + 4(B_{29}+xB_{31}+B_{32}) \right] ,
\nonumber \\[-0.00cm]
f_{1LT}(x, k_T^{\, 2}) & = - P^+ \int dk^- \left[ (A_{14}+xA_{15}) (\sigma -2x)
\right.
\nonumber \\[-0.15cm]
& \ \hspace{+1.8cm} \ 
\left.
 + A_{17} + B_{28}+xB_{30}  \right] ,
\nonumber \\[-0.05cm]
f_{1TT}(x, k_T^{\, 2}) & =  P^+\int dk^-    (A_{14}  +xA_{15} ) ,
\label{eqn:twist-2-f}
\end{align} 
where $\tau_x$ is defined by $\tau_x=\sigma^2-6x \sigma+2 \tau +6 x^2$.  
The terms of $A_{14}$, $A_{15}$, $\cdot\cdot\cdot$, and $B_{32}$
are time-reversal even (T-even) and chiral even ($\chi$-even) terms as given in
Eqs.\,(\ref{eqn:t-even-odd-terms}) and (\ref{eqn:chi-even-odd-terms}),
so that these TMDs are
T-even and $\chi$-even ones as listed in Table \ref{table:twist-2-tmd-list}.
In this table, the TMDs of the unpolarized (U), longitudinally polarized (L), 
and transversely polarized (T) hadron are also listed for showing
the complete set of the TMDs of the spin-1 hadron
in addition to the tensor polarizations $LL$, $LT$, and $TT$.
As explained after Eq.\,(\ref{eqn:spin-1-vector-tensor}),
the notation LL indicates the tensor polarization
along the longitudinal axis,
and the notations LT and TT are for 
the polarization differences shown in Appendix of Ref.\,\cite{bate2000}.

\begin{table}[b]
 \vspace{-0.30cm}
\begin{center}
   \includegraphics[width=8.5cm]{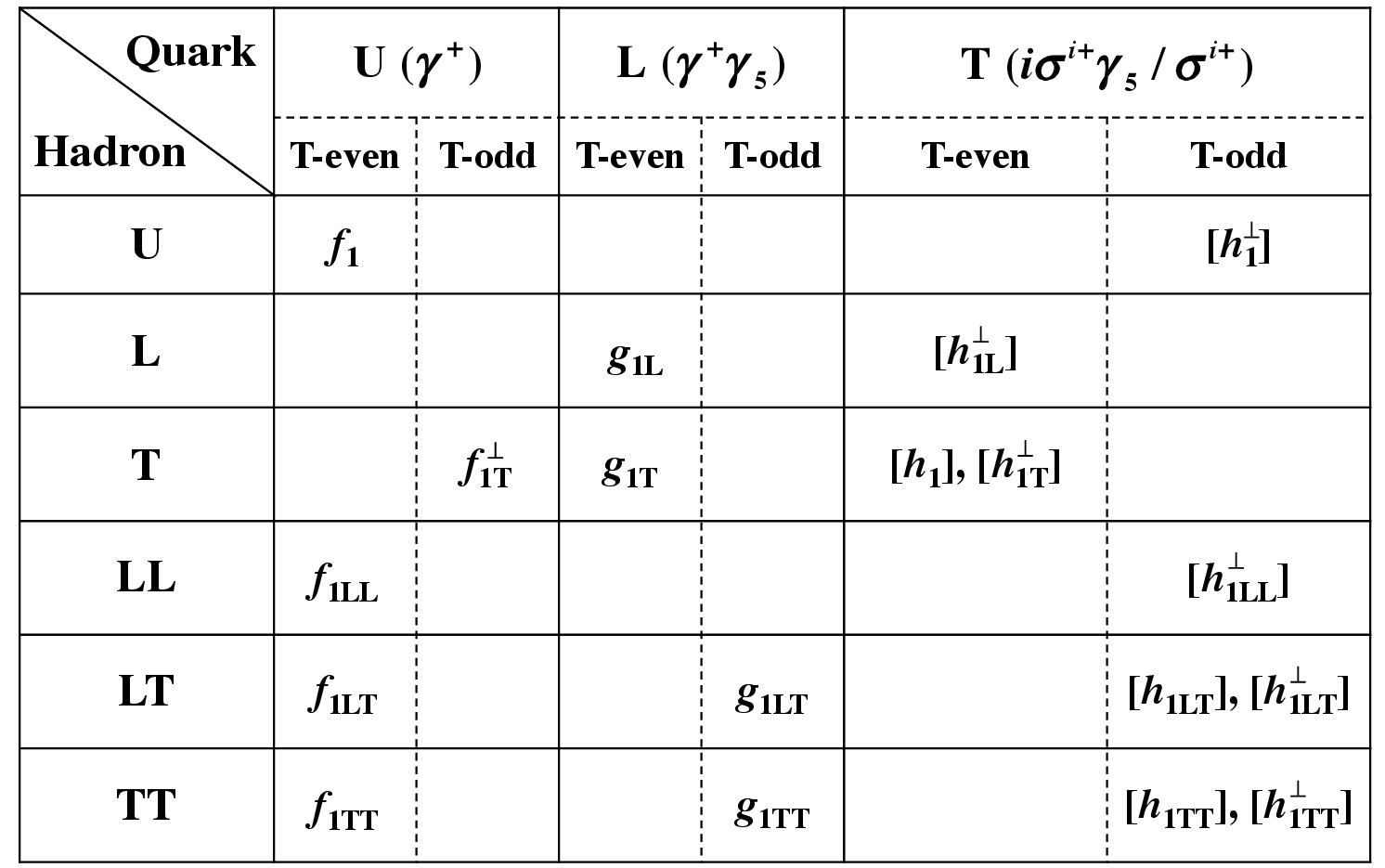}
\end{center}
\vspace{-0.5cm}
\caption{List of twist-2 quark TMDs
for a spin-1 hadron in terms of the quark and hadron polarizations.
The square brackets $[\ ]$ indicate chiral-odd distributions
and the others are chiral-even ones.
}
\label{table:twist-2-tmd-list}
\vspace{-0.00cm}
\end{table}

Next, the longitudinally polarized quark TMDs in $\Phi^{[\gamma^+ \gamma_5]}$
are given as
\begin{align}
g_{1LT}(x, k_T^{\, 2}) & =  - \frac{P^+}{2} \int dk^- 
                      \left[ A_{20} (\sigma -2x) +2B_{33} \right] ,
\nonumber \\[-0.05cm]
g_{1TT}(x, k_T^{\, 2}) & = - P^+ \int dk^- A_{20} .
\label{eqn:twist-2-g}
\end{align} 
Because of the time-reversal and chiral properties 
of the $A_{20}$ and $B_{33}$ terms,
these TMDs are T-odd and $\chi$-even distributions
as listed in Table \ref{table:twist-2-tmd-list}.
Third, 
the transversely polarized quark TMDs in 
$\Phi^{[\sigma^{i+}]}$ are given as
\begin{align}
h^{\perp}_{1LL}(x, k_T^{\, 2}) & =- \frac{P^+}{3} \int dk^-  \left[   A_{16} \tau_x  
          + 2 A_{19} (\sigma -3x) 
\right.
\nonumber \\[-0.05cm]
& \ \hspace{+0.0cm} \ 
\left.         
          + 2B_{41}  (\sigma -2x) -2(A_{18}-2 B_{42}-2B_{52}) \right] ,
\nonumber \\   
h^\prime_{1LT}(x, k_T^{\, 2}) & = 
                \frac{P^+}{2} \int dk^- 
                \left[ ( A_{18} + xA_{19}  )(\sigma -2x)  
\right.
\nonumber \\[-0.15cm]
& \ \hspace{+2.1cm} \ 
\left.                      
               + 2(B_{51}+x B_{52}) \right]  ,
\nonumber \\[-0.05cm]   
h^{\perp}_{1LT}(x, k_T^{\, 2}) & = P^+ \int dk^- \left[ A_{16} (\sigma -2x)  
               + A_{19}+ B_{41} \right] ,
\nonumber \\[-0.05cm] 
h_{1TT}^\prime(x, k_T^{\, 2}) 
& = -P^+ \int dk^- ( A_{18} + xA_{19} ) ,
\nonumber \\[-0.05cm]    
h_{1TT}^{\perp}(x, k_T^{\, 2})  & =-P^+ \int dk^-  A_{16} .
\label{eqn:twist-2-h}
\end{align} 
These TMDs are T-odd and $\chi$-odd distributions
as shown in Table \ref{table:twist-2-tmd-list}.
Here, the TMDs $h_{1LT}$ and $h_{1TT}$ are listed instead of
$h^\prime_{1LT}$ and $h^\prime_{1TT}$ due to the relation of
Eq.\,(\ref{eqn:tmd-prime-f}).
In comparison with previous works \cite{pd-drell-yan,bate2000},
the new terms exist in association with the lightcone vector $n$
and the tensor polarizations,
namely the new coefficients $B_{21 - 52}$.
Therefore, the expressions of 
$f_{1LL}$, $f_{1LT}$, $g_{1LT}$, $h^{\perp}_{1LL}$, $h^\prime_{1LT}$,
and $h^{\perp}_{1LT}$ are modified from previous ones
due to the existence of the new terms,
$B_{28-33}$, $B_{41,42}$, and $B_{51,52}$.

\begin{table}[t]
 \vspace{-0.00cm}
\begin{center}
   \includegraphics[width=8.5cm]{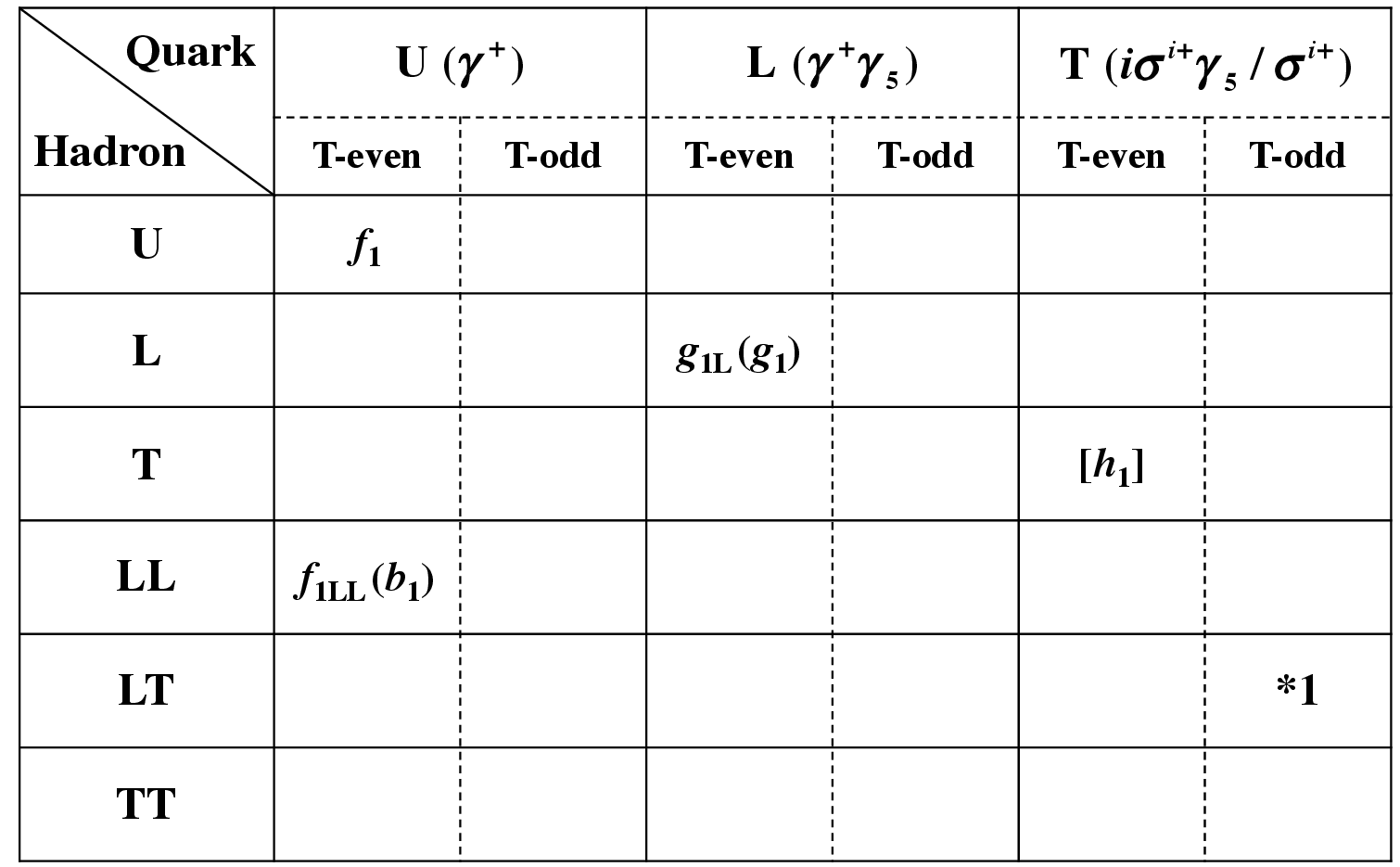}
\end{center}
\vspace{-0.5cm}
\caption{List of twist-2 quark collinear PDFs
for a spin-1 hadron in terms of the quark and hadron polarizations.
The square bracket $[\ ]$ indicates a chiral-odd distribution
and the others are chiral-even ones.
The function $g_{1L}$, $h_{1}$, and $f_{1LL}$ are often denoted as 
$g_1$ or $\Delta q$, 
$-\Delta_T q$, and
$-(2/3) b_1$ or 
$-(2/3) \delta_T q$ \cite{Kumano:2020gfk}.
\footnote{
The functions $g_{1L}$ and $f_{1LL}$ are sometimes listed
by $g_1$ and $b_1$ as for the abbreviated notations
of $g_{1,q}$ and $b_{1,q}$. These $g_1$ and $b_1$, and also
$h_1$, should not be confused with the structure function 
themselves including charge-squared factors and coefficient 
functions.
\vspace{-0.40cm}
}
The asterisk *1 is explained in the main text.
}
\label{table:twist-2-pdf-list}
\vspace{-0.00cm}
\end{table}

The collinear PDFs are obtained from the TMDs by integrating them
over the partonic transverse momentum as
\begin{align}
f (x) = \int d^2 k_T f (x, k_T^{\, 2}) .
\label{eqn:collinear-pdfs}
\end{align}
Since the time-reversal invariance is satisfied in QCD,
the T-odd collinear PDFs should vanish 
\cite{Goeke:2005hb,Metz:2008ib,Bacchetta-2007-JHEP}
\begin{align}
f (x)_{\text{T-odd}}=0 ,
\label{eqn:t-odd-pdfs}
\end{align}
although the T-odd TMDs exist in Table\,\ref{table:twist-2-tmd-list}
due to the gauge link including the transverse direction.
Therefore, the only remaining PDFs are $f_1$, $g_{1L}$ 
(or often denoted as $g_1$ or $\Delta q$), $h_1$ ($\Delta_T q$), 
and $f_{1LL}$ ($b_1$ or $\delta_T q$)
as shown in Table\,\ref{table:twist-2-pdf-list}.
The $h_1 (x)$ is defined from the TMDs $h_{1T}$ and $h_{1T}^\perp$ as
$ h_1 (x) = \int d^2 k_T \,
 [ \, h_{1T} (x,k_T^{\, 2})   
   - k_T^2 /(2M^2) \, h_{1T}^\perp (x,k_T^{\, 2}) \, ]$
\cite{bate2000}.
The number of twist-2 quark distributions is 4 in Table 3 of
Ref.\,\cite{jaffe-twist} and this number 4 agrees with 
the existence of four distributions 
$f_1$, $g_1$, $h_1$, and $f_{1LL}$ ($b_1$) 
in Table\,\ref{table:twist-2-pdf-list}. 
The only tensor-polarized twist-2 PDF is $f_{1LL}$ ($b_1$ or $\delta_T q$)
which is associated with the spin-1 nature of the hadron.
The asterisk ($*1$) in Table\,\ref{table:twist-2-pdf-list}
indicates the following.
Because of the time-reversal invariance, the collinear PDF
$h_{1LT} (x)$ vanish as shown in Eq.\,(\ref{eqn:t-odd-pdfs}).
However, since the time-reversal invariance cannot be imposed
in the fragmentation functions, we should note that
the corresponding fragmentation function $H_{1LT} (z)$ \cite{Ji-1994}, 
as indicated by the replacements of Eq.\,(\ref{eqn:tmd-fragmentation}), 
should exist as a collinear fragmentation function.

In addition to the T-odd functions, some of T-even functions
disappear after the $\vec k_T$ integration.
For example, if the correlation function 
$\Phi^{[\gamma^+]} (x, k_T, T)$ in Eq.\,(\ref{eqn:cork-2})
is integrated to obtain the collinear correlation function
\begin{align}
\Phi^{[\Gamma]} (x, T) = \int d^2 k_T \, \Phi^{[\Gamma]} (x, k_T, T) ,
\label{eqn:collinear-correlation}
\end{align}
where $\Gamma=\gamma^+$ here,
the second term vanishes and the third term also vanishes 
due to $S_{TT}^{xx}=-S_{TT}^{yy}$
\cite{Kumano:2020gfk},
so that the collinear PDFs $f_{1LT} (x)$ and $f_{1TT} (x)$
do not exist. 
In the same way, the functions
$g_{1LT}$, $g_{1TT}$, and $h_{1L}$
do not exist in Table \ref{table:twist-2-pdf-list}.

\vspace{-0.20cm}

\begin{widetext}
\subsection{Twist-3 TMDs for a tensor-polarized spin-1 hadron}
\label{twist-3}

For the spin-1/2 nucleon, the twist-3 TMDs are listed in 
the quark correlation function
including new terms with the lightcone vector $n$  
in Ref.\,\cite{Goeke:2005hb}.
Here, we list all the possible twist-3 TMDs
in the quark correlation function for a tensor-polarized spin-1 hadron,
so that all the following terms are new ones we found
in this work.
The twist-3 TMDs with the $1/P^+$ dependence 
are found by considering 
$\Phi^{ [ \gamma^i ] }$,
$\Phi^{\left[{\bf 1}\right]}$,
$\Phi^{\left[i\gamma_5\right]}$
$\Phi^{ [\gamma^{i}\gamma_5 ]}$
$\Phi^{ [ \sigma^{ij} ]}$,
and $\Phi^{ [ \sigma^{-+} ] }$.
First, the TMDs with the function name $f$ are defined for
the quark operator type $\bar\psi \gamma^i \psi$ as
\begin{align}
\Phi^{ [ \gamma^i ] } (x, k_T, T)
= & 
\frac{M}{P^+} \left[  f^{\perp}_{LL}(x, k_T^{\, 2})  S_{LL} \frac{k_T^i}{M}
+ f^{\,\prime} _{LT} (x, k_T^{\, 2})S_{LT}^i 
- f_{LT}^{\perp}(x, k_T^{\, 2}) \frac{ k_{T}^i  S_{LT}\cdot k_{T}}{M^2} 
- f_{TT}^{\,\prime} (x, k_T^{\, 2}) \frac{S_{TT}^{ i j} k_{T \, j} }{M} 
\right. 
\nonumber \\[-0.05cm]
& \ \ \ \ \ 
\left. 
+ f_{TT}^{\perp}(x, k_T^{\, 2}) \frac{k_T\cdot S_{TT}\cdot k_T}{M^2} 
       \frac{k_T^i}{M} \right] .
\label{eqn:cork-3-1a}
\end{align} 
These $f$-type TMDs have T-even and $\chi$-even properties.
The distributions $f_{LT}$, 
$f^\prime_{LT}$, and $f^\perp_{LT}$
($f_{TT}$, $f^\prime_{TT}$, and $f^\perp_{TT}$)
are related by the relation of Eq.\,(\ref{eqn:tmd-prime-f}).
The TMDs with the name $e$ are assigned for the currents associated with
$\bar\psi {\bf 1} \psi$ and $\bar\psi i \gamma_5 \psi$ as
\begin{align}
\Phi^{\left[ {\bf 1} \right]} (x, k_T, T)
= &\frac{M}{P^+} \left[ e_{LL}(x, k_T^{\, 2}) S_{LL}
- e_{LT}^\perp (x, k_T^{\, 2}) \frac{S_{LT}\cdot k_T}{M} 
+ e_{TT}^\perp (x, k_T^{\, 2})  \frac{k_T\cdot S_{TT}\cdot k_T}{M^2}  \right] ,
\nonumber \\
\Phi^{\left[i\gamma_5\right]}= & 
     \frac{M}{P^+} \left[ e_{LT} (x, k_T^{\, 2}) \frac{S_{LT\, \mu}
                                 \varepsilon_{T}^{\mu\nu} k_{T\, \nu}}{M}
    -e_{TT} (x, k_T^{\, 2}) \frac{S_{TT\, \mu \rho} k_{T}^{\rho} 
   \varepsilon_{T}^{\mu\nu} k_{T\, \nu}}{M^2}  \right] .
\label{eqn:cork-3-1b}
\end{align} 
These $e$-type TMDs have T-even and $\chi$-odd properties.
The distributions $e_{LT}$ and $e_{TL}^\perp$ 
are given with the same factors of $O((k_T)^1)$; 
however, we assigned $e_{LT}^\perp$ 
for the first one and $e_{LT}$ for the second
as explained in the guideline 6 of Sec.\,\ref{tmd-name-gauideline}.
The distributions $e_{TT}$ and $e_{TT}^\perp$ are also named
in the same way.
Next, the $g$-type TMDs with T-odd and $\chi$-even properties
are defined for the current $\bar\psi \gamma^{i}\gamma_5 \psi$ as
\begin{align}
\Phi^{ [\gamma^{i}\gamma_5 ]} (x, k_T, T)
= & \frac{M}{P^+} 
       \left[  
      - g^{\perp}_{LL}(x, k_T^{\, 2})  
       S_{LL}  \frac{  \varepsilon_{T}^{ij} k_{T\, j}}{M}
      - g^{\,\prime}_{LT} (x, k_T^{\, 2}) \varepsilon_{T}^{ij}  S_{LT\,j} 
+ g_{LT}^{\perp}(x, k_T^{\, 2}) \frac{  \varepsilon_{T}^{ij} k_{T\, j}
               S_{LT}\cdot k_{T}}{M^2} 
\right.
\nonumber \\[-0.05cm]
& \ \ \ \ \ 
\left. 
+ g_{TT}^{\,\prime} (x, k_T^{\, 2})  \frac{  \varepsilon_{T}^{ij} 
               S_{TT\, jl}k_{T}^l }{M} 
- g_{TT}^{\perp}(x, k_T^{\, 2}) \frac{k_T\cdot S_{TT}\cdot k_T}{M^2} 
               \frac{ \varepsilon_{T}^{ij} k_{T\, j}}{M} \right] .
\label{eqn:cork-3-2}
\end{align} 
The distributions $g_{LT}$, $g^\prime_{LT}$, and $g^\perp_{LT}$
($g_{TT}$, $g^\prime_{TT}$, and $g^\perp_{TT}$)
are related by the relation of Eq.\,(\ref{eqn:tmd-prime-f}).
The $h$-type TMDs with the T-odd and $\chi$-odd properties
are given for the currents
$\bar\psi \sigma^{-+} \psi$ and $\bar\psi \sigma^{ij} \psi$ as
\begin{align}
\Phi^{ [ \sigma^{-+} ] } (x, k_T, T)
= &
\frac{M}{P^+} \left[ h_{LL}(x, k_T^{\, 2}) S_{LL}
- h_{LT}(x, k_T^{\, 2}) \frac{S_{LT}\cdot k_T}{M} 
+ h_{TT}(x, k_T^{\, 2}) \frac{k_T\cdot S_{TT}\cdot k_T}{M^2} \right] ,
\nonumber \\
\Phi^{ [ \sigma^{ij} ]} (x, k_T, T)
= &
\frac{M}{P^+} \left[ h_{LT}^{\perp}(x, k_T^{\, 2}) 
              \frac{S_{LT}^i k_T^j-  S_{LT}^j k_T^i}{M} 
- h_{TT}^{\perp}(x, k_T^{\, 2}) 
\frac{ S_{TT}^{il} k_{T\, l}  k_T^j 
- S_{TT}^{jl} k_{T\, l}  k_T^i   }{M^2} \right] .
\label{eqn:cork-3-3}
\end{align} 
The prime marks ($\prime$) are not assigned for 
$h_{LT}$ and $h_{TT}$ because of the guideline 6 
in Sec.\ref{tmd-name-gauideline}.
\end{widetext}

There are twenty TMDs in the twist-3 for a tensor-polarized spin-1 hadron.
These TMDs are expressed by the expansion coefficients
of the correlation function.
First, we obtain the $f$-type TMDs as
\begin{align}
f^{\perp}_{LL}(x, k_T^{\, 2}) & =  \frac{P^+} {3}\int dk^-  
       \left[  A_{15} \tau_x +2A_{17}     \right.
\nonumber \\[-0.15cm]
& \ \hspace{2.0cm} \        
       \left.
       +2B_{30}(\sigma-2x) +4 B_{31} \right] ,
\nonumber \\[-0.05cm]
f^{\,\prime}_{LT} (x, k_T^{\, 2}) & =\frac{P^+} {2}\int dk^- 
                        \left[  A_{17} (\sigma-2x) +2 B_{32} \right] ,
\nonumber \\[-0.05cm] 
f^{\perp}_{LT}(x, k_T^{\, 2}) & =
- P^+ \int dk^-  \left[ A_{15} (\sigma-2x) + B_{30} \right] ,
\nonumber \\[-0.05cm]  
f_{TT}^{\,\prime} (x, k_T^{\, 2}) & = - P^+ \int dk^-  A_{17} ,
\nonumber \\[-0.05cm]  
f_{TT}^{\perp}(x, k_T^{\, 2}) & = P^+\int dk^-  A_{15} .
\label{eqn:twist-3-f}
\end{align} 
The terms with $A_{15}$, $A_{17}$, $\cdots$, $B_{32}$ are
T-even and $\chi$-even as listed in 
Eqs.\,(\ref{eqn:t-even-odd-terms}) and (\ref{eqn:chi-even-odd-terms}),
so that these TMDs are T-even and $\chi$-even properties as shown 
in Table\,\ref{table:twist-3-tmd-list}.
The $e$-type TMDs are expressed as
\begin{align}
e_{LL}(x, k_T^{\, 2}) & = \frac{P^+} {3}\int dk^- 
         \left[  A_{13} \tau_x +2B_{21}(\sigma-2x) +4 B_{22} \right] ,    
\nonumber \\[-0.05cm] 
e_{LT}^\perp (x, k_T^{\, 2}) & = - P^+
\int dk^-   \left[ A_{13}(\sigma-2x) +B_{21} \right] ,
\nonumber \\[-0.05cm]
e_{TT}^\perp (x, k_T^{\, 2}) & = P^+\int dk^-  A_{13} ,
\nonumber \\[-0.05cm]
e_{LT} (x, k_T^{\, 2}) & = P^+\int dk^-  
                  \left[ B_{23}(x- \frac{\sigma}{2}) - B_{24} \right] ,
\nonumber \\[-0.05cm] 
e_{TT} (x, k_T^{\, 2}) & = P^+ \int dk^-  B_{23} .
\label{eqn:twist-3-e}
\end{align} 
Because of the terms $A_{13}$, $B_{21}$, $\cdots$, $B_{24}$,
these TMDs have properties of T-even and $\chi$-odd
as shown in Table\,\ref{table:twist-3-tmd-list}.

Second, we obtain the $g$-type TMDs as
\begin{align}
g^{\perp}_{LL}(x, k_T^{\, 2}) & = 
 \frac{P^+} {3} \int dk^- \left[ 3 A_{20} (\sigma-2x) + 4B_{33}+2B_{34}
\right.
\nonumber \\[-0.15cm]
& \ \hspace{+1.7cm} \ 
\left.                              
        +B_{36}\tau_x+2 B_{37  } (\sigma-2x)+ 4B_{38} \right]  ,
\nonumber \\[-0.05cm]  
g^{\,\prime}_{LT} (x, k_T^{\, 2}) & =  \frac{P^+} {4}
     \int dk^-  \left[  A_{20} (\sigma-2x)^2 
\right.
\nonumber \\[-0.15cm]
& \ \hspace{+1.7cm} \ 
\left.                              
        + 2(B_{33}+B_{34})(\sigma-2x)+ 4B_{35} \right] ,
\nonumber \\   
g_{LT}^{\perp}(x, k_T^{\, 2}) & = - P^+\int dk^-
                  \left[A_{20} +B_{36}(\sigma-2x)+B_{37} \right] ,
\nonumber \\[-0.05cm]       
g_{TT}^{\,\prime} (x, k_T^{\, 2}) & = -P^+\int dk^- 
                   \left[A_{20} (\frac{\sigma}{2}-x)+B_{34} \right] ,
\nonumber \\[-0.05cm]  
g_{TT}^{\perp}(x, k_T^{\, 2}) & = P^+ \int dk^- B_{36} .
\label{eqn:twist-3-g}
\end{align} 
Because of the terms $A_{20}$, $\cdots$, $B_{38}$,
these TMDs have properties of T-odd and $\chi$-even
as shown in Table\,\ref{table:twist-3-tmd-list}.
Three new twist-3 TMDs $e_{LT}$, $e_{TT}$, and $g_{TT}^{\perp}$ 
are expressed purely by the new terms $B_{20-52}$.

\begin{table}[t]
 \vspace{-0.00cm}
\begin{center}
   \includegraphics[width=8.5cm]{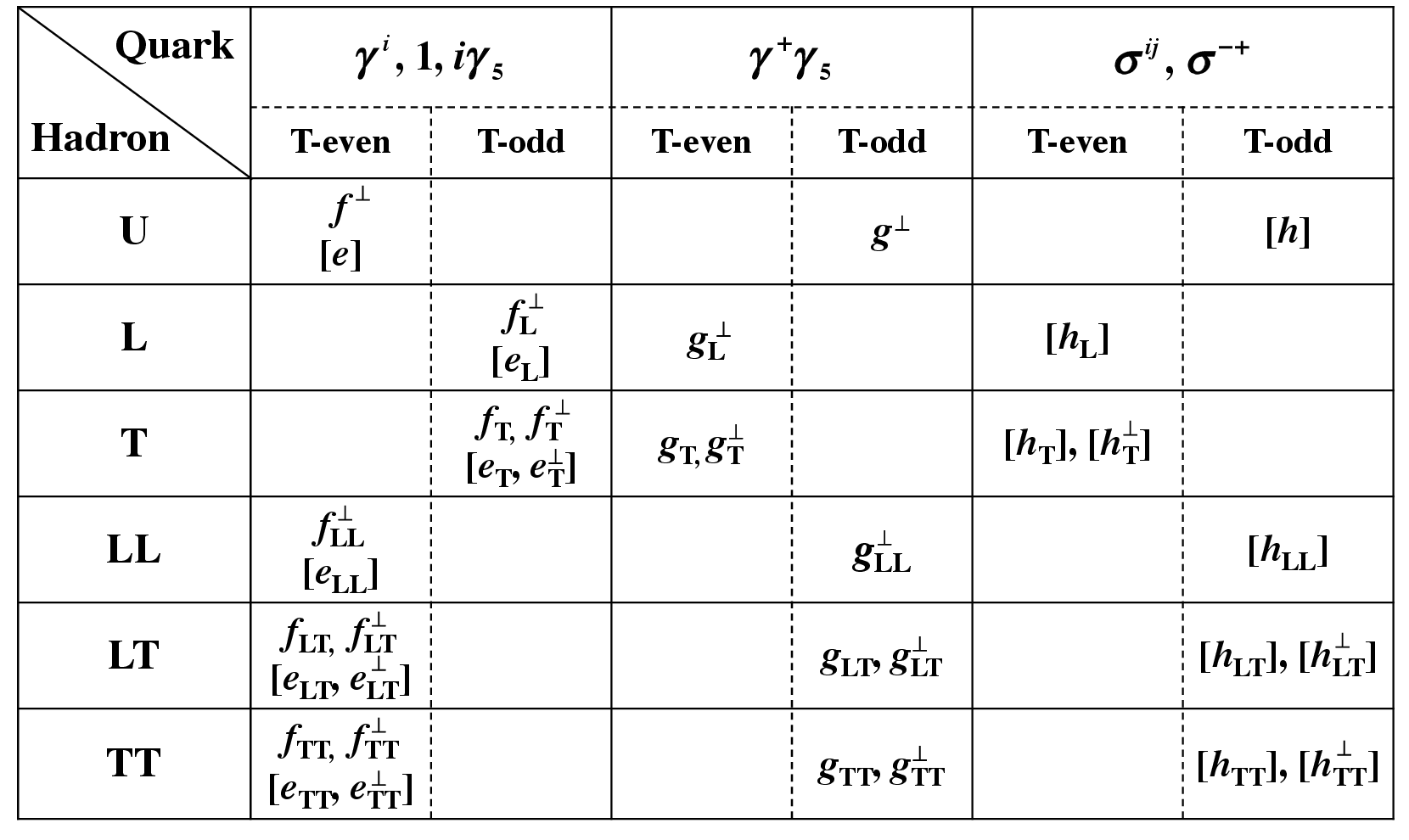}
\end{center}
\vspace{-0.4cm}
\caption{List of twist-3 quark TMDs
for a spin-1 hadron in terms of the hadron polarizations
and the operator forms in the correlation functions.
The square brackets $[\ ]$ indicate chiral-odd distributions
and the others are chiral-even ones.
The LL, LT, and TT TMDs are new distributions found in this work.
}
\label{table:twist-3-tmd-list}
\vspace{-0.30cm}
\end{table}

\begin{table}[b]
 \vspace{-0.00cm}
\begin{center}
   \includegraphics[width=8.5cm]{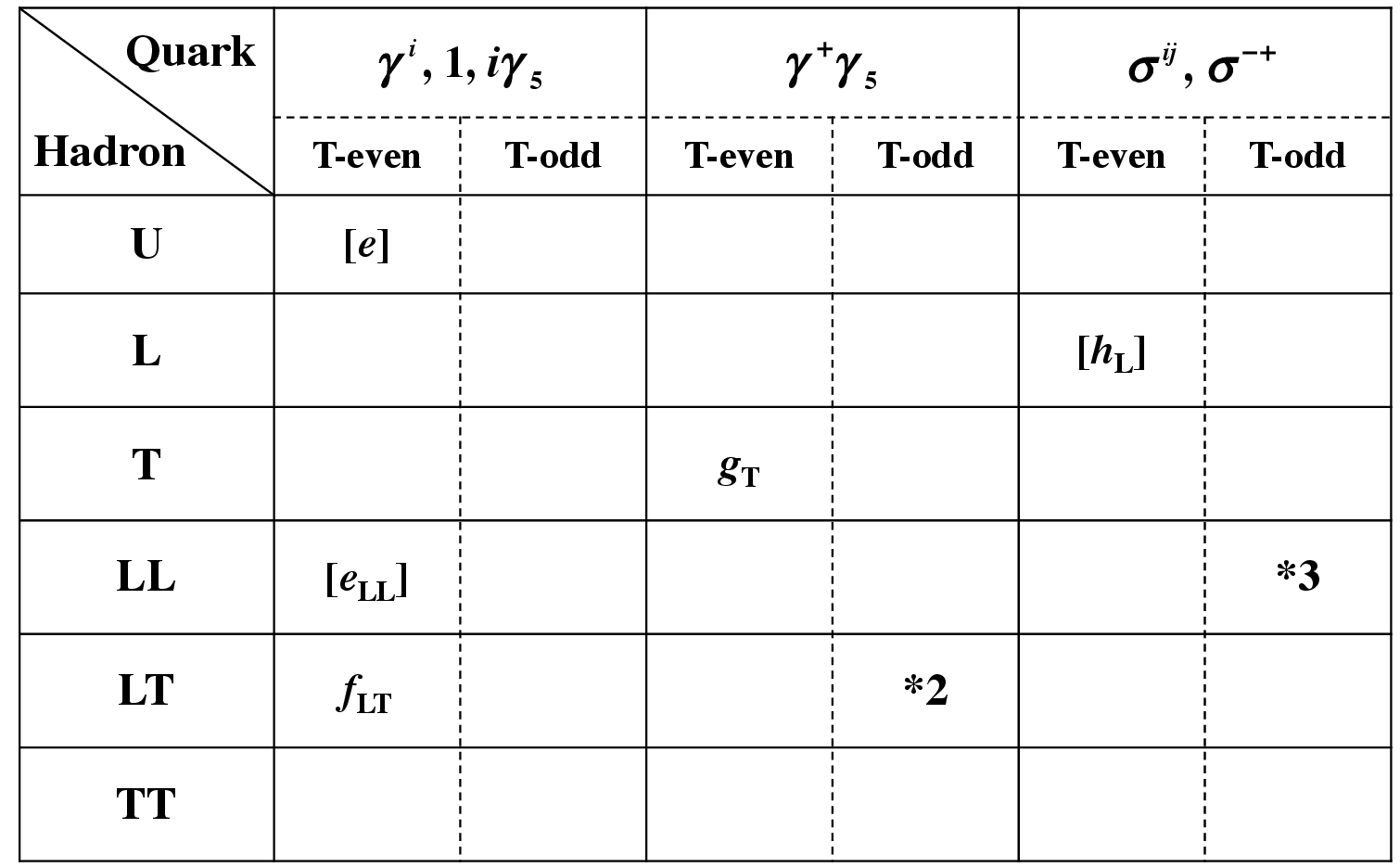}
\end{center}
\vspace{-0.4cm}
\caption{List of twist-3 quark collinear PDFs
for a spin-1 hadron in terms of the hadron polarizations
and the operator forms in the correlation functions.
The square brackets $[\ ]$ indicate a chiral-odd distribution
and the others are chiral-even ones.
The LL and LT PDFs ($e_{LL}$, $f_{LT}$) are new distributions 
found in this work.
The asterisks *2 and *3 are explained in the main text.
}
\label{table:twist-3-pdf-list}
\vspace{+0.30cm}
\end{table}

Third, the $h$-type TMDs are expressed as
\begin{align}
\displaybreak[1]
h_{LL}(x, k_T^{\, 2}) & = \frac{P^+} {6}\int dk^-  
                    \left \{  (\sigma-2x) \left[A_{16} \tau_x 
+2 (2 A_{18}+ \sigma A_{19})  \right]  
\right.
\nonumber \\[-0.05cm]
& \ \hspace{-2.0cm} \ 
\left.                              
+2(\sigma-2x) \left[ B_{41}(\sigma-2x) 
+2(B_{42}+B_{44}+x B_{47}+ B_{49} )  \right]     
\right.
\nonumber \\[-0.05cm]
& \ \hspace{-2.0cm} \ 
\left.                              
 +2(B_{43}+xB_{46}) \tau_x   +4(2 B_{45}+ 2x B_{48}+ 2B_{50}+ 2 B_{51}
          + \sigma B_{52})  \right\} ,
\nonumber \\[-0.05cm]
h_{LT}(x, k_T^{\, 2}) & = - \frac{P^+} {2}
           \int dk^- \left[  A_{16} (\sigma-2x)^2  
          + 2A_{18}+\sigma A_{19}  
\right.
\nonumber \\[-0.05cm]
& \ \hspace{-2.0cm} \ 
\left.                                        
          +(\sigma-2x) (  B_{41}+ 2 B_{43}+ 2x B_{46}) 
          + 2(B_{44}+x B_{47}+ B_{49} )  \right] ,
\nonumber \\[-0.05cm]       
h_{LT}^{\perp}(x, k_T^{\, 2}) & =\frac{P^+} {2}\int dk^-
                          \left[  A_{19} (\sigma-2x) +2 B_{52} \right] ,
\nonumber \\[-0.05cm]
h_{TT}(x, k_T^{\, 2}) & =\frac{P^+} {2}\int dk^- \left[ A_{16} (\sigma-2x) 
           +2(B_{43}+ xB_{46}) \right] ,
\nonumber \\[-0.05cm]
h_{TT}^{\perp}(x, k_T^{\, 2}) & = - P^+ \int dk^-  A_{19} .
\label{eqn:twist-3-h}
\end{align}  
\vfill\eject\noindent
Because of the terms $A_{16}$, $\cdots$, $B_{52}$,
these TMDs have properties of T-odd and $\chi$-odd
as shown in Table\,\ref{table:twist-3-tmd-list}.
Here, the listed functions $f_{LT}$, $f_{TT}$, 
$h_{LT}$, and $h_{TT}$ are defined from
$f^\prime_{LT}$, $f^\prime_{TT}$, $h^\prime_{LT}$, and $h^\prime_{TT}$, 
and 
$f_{LT}^\perp$, $f_{TT}^\perp$, $h_{LT}^\perp$, and $h_{TT}^\perp$ 
by the relation in Eq.\,(\ref{eqn:tmd-prime-f}).

These TMDs are integrated over the quark transverse momentum $\vec k_T$,
twist-3 collinear PDFs exist as shown in Table \ref{table:twist-3-pdf-list}.
In addition to the PDFs $e$, $g_T$, and $h_L$ in the nucleon,
there are new twist-3 PDFs $e_{LL}$ and $f_{LT}$ 
for the spin-1 hadron.
These two collinear PDFs are new functions found in this work
by integrating the corresponding TMDs over $\vec k_T$.
Here, the PDF $g_T$ is given by 
$ g_T = \int d^2k_T [ g_T^{\,\prime} 
- k_T^2 /(2 M^2) g_T^{\perp} ] $
\cite{Goeke:2005hb},
and $f_{LT} (x)$ is defined by
\begin{align}
f_{LT} (x) & = \int d^2k_T \, f_{LT} (x, k_T^{\, 2}) .
\label{eqn:f-lt}
\end{align}
The asterisks ($*2$ and $*3$) in Table\,\ref{table:twist-3-pdf-list}
indicate the following in the same way with $*1$.
Because of the time-reversal invariance, the collinear PDFs
$g_{LT} (x)$ and $h_{LL} (x)$ do not exist. 
However, the corresponding new collinear
fragmentation functions $G_{LT} (z)$ and $H_{LL} (z)$
should exist \cite{Ji-1994}.

\vspace{-0.00cm}

\begin{widetext}
\subsection{Twist-4 TMDs for a tensor-polarized spin-1 hadron}
\label{twist-4}

The twist-4 TMDs were obtained in Ref.\,\cite{Goeke:2005hb}
for the spin-1/2 nucleon. Here, we list all the possible twist-4 quark
TMDs for the tensor-polarized spin-1 hadron.
The twist-4 TMDs for the tensor-polarized spin-1 hadron are defined 
in the correlation functions $\Phi^{[\gamma^-]}$,
$\Phi^{[\gamma^- \gamma_5]}$, and $\Phi^{[\sigma^{i-}]}$ as
\begin{align}
\Phi^{[\gamma^-]}= & \frac{M^2}{P^{+2}} \left[ f_{3LL}(x, k_T^{\, 2}) S_{LL}
- f_{3LT}(x, k_T^{\, 2}) \frac{S_{LT}\cdot k_T}{M} 
+ f_{3TT}(x, k_T^{\, 2}) \frac{k_T\cdot S_{TT}\cdot k_T}{M^2} \right] ,
\nonumber \\
\Phi^{[\gamma^- \gamma_5]}= & 
\frac{M^2}{P^{+2}} \left[ g_{3LT}(x, k_T^{\, 2}) \frac{S_{LT\, \mu} 
                            \varepsilon_{T}^{\mu\nu} k_{T\, \nu}}{M} 
+ g_{3TT}(x, k_T^{\, 2})  \frac{S_{TT\, \mu \rho} k_{T}^{\rho} 
                           \varepsilon_{T}^{\mu\nu} k_{T\, \nu}}{M^2} \right] ,
\nonumber \\
\Phi^{[\sigma^{i-}]}= & 
\frac{M^2}{P^{+2}} \left[  h^{\perp}_{3LL}(x, k_T^{\, 2})S_{LL} \frac{k_T^i}{M}
+ h^{'}_{3LT}(x, k_T^{\, 2})S_{LT}^i 
-  h_{3LT}^{\perp}(x, k_T^{\, 2}) \frac{ k_{T}^i S_{LT}\cdot k_{T}}{M^2}
- h_{3TT}^{\prime} (x, k_T^{\, 2}) \frac{S_{TT}^{ i j} k_{T \, j} }{M} \right.
\nonumber \\[-0.15cm]
& \left. 
+ h_{3TT}^{\perp}(x, k_T^{\, 2}) \frac{k_T\cdot S_{TT}\cdot k_T}{M^2}
         \frac{k_T^i}{M} \right] .
\label{eqn:cork-4}
\end{align} 
These relations are proportional to $1/(P^+)^2$
as the twist-4 functions.
\vspace{-0.10cm}
\end{widetext}

\ \ 

\vfill\eject

\ \ 

\vfill\eject

The $f$-type TMDs are given by
\begin{align}
\! \! \! \! \! \! \!
f_{3LL}(x, k_T^{\, 2}) & = 
      \frac{P^+} {6} \! \int \! dk^- \left \{ \left[A_{14} 
      +A_{15}(\sigma-x) \right] \tau_x  
\right.
\nonumber \\[-0.05cm]
& \ \hspace{-0.0cm} \ 
\left.                              
       - 2 A_{17}(\sigma-2x) + 2 B_{25}\tau_x
\right.
\nonumber \\[-0.05cm]
& \ \hspace{-0.0cm} \ \left.                              
       +2(\sigma-2x) \left[ 2B_{26}+B_{28}+ B_{30}(\sigma-x)) \right]  
\right.   
\nonumber \\[-0.05cm] 
& \ \hspace{-0.0cm} \ \left.   
      + 4(2B_{27}+B_{29}- B_{32}) +4B_{31}(\sigma-x)   \right\} ,
\nonumber \\[-0.05cm]
\! \! \! \! \! \! \!
f_{3LT}(x, k_T^{\, 2}) & = - \frac{P^+} {2} 
      \! \int \! dk^- \left \{ \left[A_{14} 
                    +A_{15}(\sigma-x) \right] (\sigma-2x)
\right.   
\nonumber \\[-0.05cm]
& \ \hspace{-1.5cm} \ \left.   
   -A_{17}  +2B_{25}(\sigma-2x) + 2B_{26}+B_{28} +B_{30}(\sigma-x) \right \} ,
\nonumber \\ 
\! \! \! \! \! \! \!
f_{3TT}(x, k_T^{\, 2}) & = \frac{P^+} {2} \! \int \! dk^- \left[A_{14} +A_{15}(\sigma-x) 
                       + 2 B_{25} \right] .
\label{eqn:twist-4-f}
\end{align} 
Because of the terms $A_{14}$, $\cdots$, $B_{25}$,
these TMDs have properties of T-even and $\chi$-even
as shown in Table\,\ref{table:twist-4-tmd-list}.
The $g$-type TMDs are
\begin{align}
g_{3LT}(x, k_T^{\, 2}) & = \frac{P^+} {4}\int dk^- 
            \left[(A_{20} -2B_{39})(\sigma-2x) 
\right.
\nonumber \\[-0.15cm]
& \ \hspace{+2.0cm} \ 
\left.                                          
            + 2 B_{33} -4 B_{40} \right] ,
\nonumber \\[-0.05cm]
g_{3TT}(x, k_T^{\, 2}) & = \frac{P^+} {2}\int dk^-   ( A_{20} -2B_{39}) .
\label{eqn:twist-4-g}
\end{align} 
These TMDs have the properties of 
T-odd and $\chi$-even as shown in Table\,\ref{table:twist-4-tmd-list}.
The $h$-type TMDs are 
\begin{align}
h_{3LL}^{\perp}(x, k_T^{\, 2}) & = \frac{P^+} {6}\int dk^- 
               \left[ -A_{16}  \tau_x + 2 A_{18} 
\right.   
\nonumber \\[-0.05cm]
& \ \hspace{0.0cm} \ \left.   
         + 2 A_{19}(2\sigma-3x) - 2( B_{41}-2B_{47}) (\sigma-2x) 
\right.   
\nonumber \\[-0.05cm]
& \ \hspace{0.0cm} \ \left.     
         + 2B_{46} \tau_x - 4(B_{42}- 2B_{48} -B_{49}-  B_{52})  \right] ,
\nonumber \\[-0.05cm]  
h^\prime_{3LT}(x, k_T^{\, 2}) & = \frac{P^+} {4} \int dk^- 
        \left\{ \left[  A_{18}+  A_{19} (\sigma-x)  \right] (\sigma-2x) 
\right.
\nonumber \\[-0.00cm]
& \ \hspace{-1.0cm} \ 
\left.                                          
        + 2\left[B_{49}(\sigma-2x) 
        +2B_{50} +B_{51} + B_{52} (\sigma-x) \right] \right\} ,
\nonumber \\ 
h^{\perp}_{3LT}(x, k_T^{\, 2}) & = - \frac{P^+} {2}
\int dk^- \left[ -A_{16}(\sigma-2x) +A_{19} -B_{41} 
\right.   
\nonumber \\[-0.05cm]
& \ \hspace{2.0cm} \ \left.     
              + 2 B_{46} (\sigma-2x) +2B_{47}  \right] ,
\nonumber \\[-0.05cm] 
h^\prime_{3TT} (x, k_T^{\, 2}) & =
- \frac{P^+} {2} \int dk^-  \left[ A_{18}+ A_{19}(\sigma-x) + 2 B_{49} \right] ,
\nonumber \\[-0.05cm] 
h^{\perp }_{3TT}(x, k_T^{\, 2}) & =-\frac{P^+} {2} \int dk^- 
                 \left[ A_{16}-2 B_{46} \right] .
\label{eqn:twist-4-h}
\end{align}    
\begin{table}[t]
 \vspace{-0.00cm}
\begin{center}
   \includegraphics[width=8.5cm]{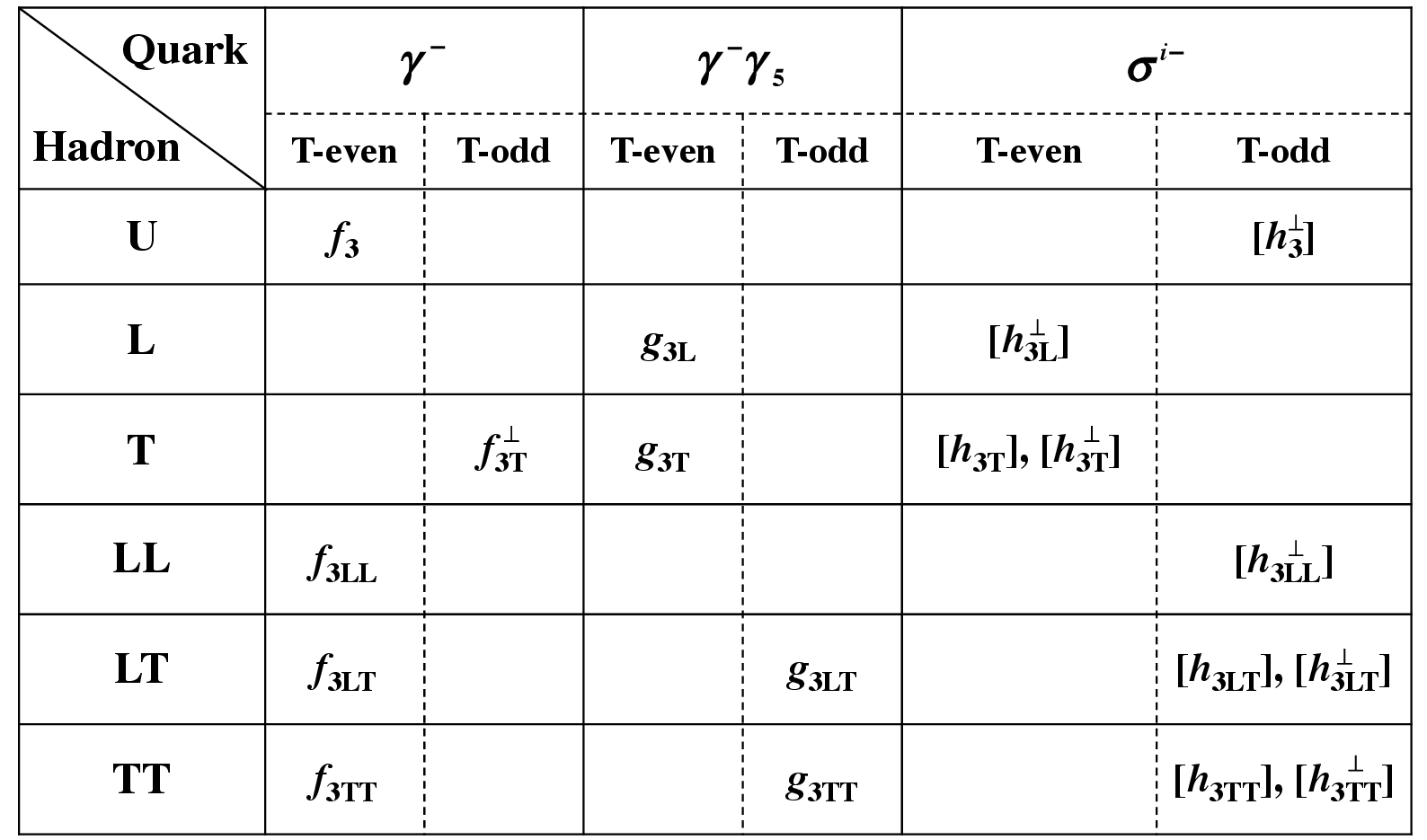}
\end{center}
\vspace{-0.4cm}
\caption{List of twist-4 quark TMDs
for a spin-1 hadron in terms of the hadron polarizations
and the operator forms in the correlation functions.
The square brackets $[\ ]$ indicate chiral-odd distributions
and the others are chiral-even ones.
The LL, LT, and TT TMDs are new distributions found 
in this work.}
\label{table:twist-4-tmd-list}
\vspace{-0.00cm}
\end{table}
\begin{table}[t]
 \vspace{-0.00cm}
\begin{center}
   \includegraphics[width=8.1cm]{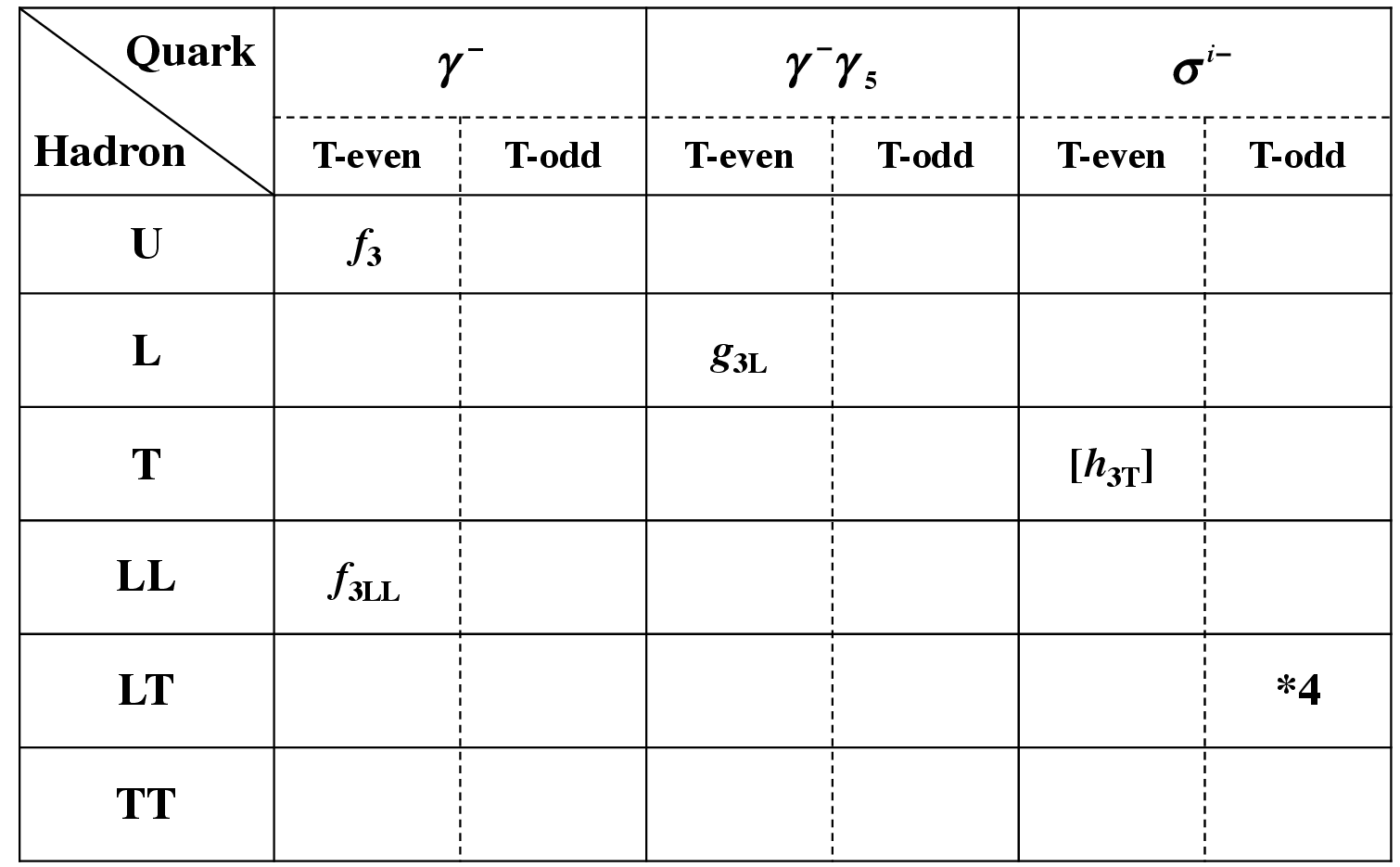}
\end{center}
\vspace{-0.4cm}
\caption{List of twist-4 quark collinear PDFs
for a spin-1 hadron in terms of the hadron polarizations
and the operator forms in the correlation functions.
The square bracket $[\ ]$ indicates a chiral-odd distribution
and the others are chiral-even ones.
The LL PDF ($f_{3LL}$) is a new distribution found in this work.
The asterisk *4 is explained in the main text.
}
\label{table:twist-4-pdf-list}
\vspace{-0.30cm}
\end{table}
These TMDs have the properties of 
T-odd and $\chi$-odd
as shown in Table\,\ref{table:twist-4-tmd-list}.
Here, the functions $h_{3T}$, $h_{3LT}$, and $h_{3TT}$ 
are defined from $h^\prime_{3T}$, $h^\prime_{3LT}$, and $h^\prime_{3TT}$
and $h_{3T}^\perp$, $h_{3LT}^\perp$, and $h_{3TT}^\perp$
by the relation in Eq.\,(\ref{eqn:tmd-prime-f}).

These twist-4 TMDs are integrated over $\vec k_T$ and they
become the collinear PDFs as shown in Table\,\ref{table:twist-4-pdf-list}.
Most distributions vanish after the integrations. The only
twist-4 PDF which is specific to the tensor-polarized spin-1 hadron
is $f_{3LL}$, in addition to $f_3$, $g_{3L}$, and $h_{3L}$
which exist also for the spin-1/2 nucleon.
The asterisk ($*4$) in Table\,\ref{table:twist-4-pdf-list}
indicates that $h_{3LT} (x)$ does not exist; however,
the corresponding new collinear fragmentation function $H_{3LT} (z)$
should exist because the time-reversal invariance does not have 
to be imposed \cite{Ji-1994}.

\subsection{Summary on new TMDs and possible new fragmentation functions}
\label{comments-tmds-ffs}

We found that there are 40 TMDs in total for the tensor-polarized spin-1
hadron, and this number is equal to the one of the expansion terms 
in Eq.\,(\ref{eqn:cork4}), and they are expressed by 
the coefficients $A_i$ and $B_i$.
The TMDs are T-odd if they are associated with the gamma matrices 
$\gamma^\mu \gamma_5$ and $\sigma^{\mu \nu}$ in the tensor-polarized case, 
so that there are 24 T-odd TMDs.
In addition, there are 16 T-even TMDs on the tensor polarizations. 
If the gauge link were neglected in the correlation function, 
all the T-odd TMDs do not exist due to the time-reversal invariance.
The 10 twist-2 TMDs were studied in Ref.\,\cite{bate2000}, 
so we found 30 new TMDs in the twist-3 and 4 parts mainly associated with 
the lightcone vector $n$ and the tensor polarizations
as listed in Tables \ref{table:twist-3-tmd-list}
and \ref{table:twist-4-tmd-list}.

The same discussions can be made for 
the transverse-momentum-dependent fragmentation functions of spin-1 hadrons 
by the replacements of the kinematical variables 
and the function notations as 
\cite{bate2000}
\begin{align}
& \text{Kinematical variables:}   \ \  x,\, k_T,\, S,\, T,\, M,\, n ,
\, \gamma^+,\, \sigma^{i+},
\nonumber \\
& \text{TMD distribution functions:}  \ \ f,\, g,\, h,\, e
\nonumber \\
& \Downarrow    \  \ 
\nonumber \\
& \text{Kinematical variables:}   \ \  z,\, k_T,\, S_h,\, T_h,\, M_h,\, 
\bar n ,
\, \gamma^-,\, \sigma^{i-},
\nonumber \\
& \text{TMD fragmentation functions:}  \ \  D,\, G,\, H,\, E .
\label{eqn:tmd-fragmentation}
\end{align} 
Therefore, new fragmentation functions exist for spin-1 hadrons
in addition to the fragmentation functions of the spin-1/2 nucleon
by these simple replacements in Tables
\ref{table:twist-2-tmd-list},
\ref{table:twist-2-pdf-list},
\ref{table:twist-3-tmd-list},
\ref{table:twist-3-pdf-list},
\ref{table:twist-4-tmd-list}, and
\ref{table:twist-4-pdf-list}.
Here, $S_h$ and $T_h$ are spin-vector and tensor polarizations
of the hadron $h$, and $M_h$ is its mass.
The variable $z$ is the momentum fraction given by
$P_h^- = z k^- $.
As explained by the asterisks ($*1$--$4$) in the collinear PDF tables,
there are the collinear fragmentation functions $H_{1LT} (z)$, 
$G_{LT}(z)$, $H_{LL}(z)$, and $H_{3LT} (z)$ 
although their corresponding functions 
$h_{1LT} (x)$, $g_{LT}(x)$, $h_{LL}(x)$, and $h_{3LT} (x)$
vanish due to the time-reversal invariance.

\subsection{Integral relations in T-odd TMDs}
\label{t-odd-sum}

If we integrate the $k_T$-dependent correlation function $\Phi(x, k_T, T)$
over $k_T$, the T-odd terms should vanish on account 
of time-reversal invariance ($\int d^2 k_T \, \Phi_{\text{T-odd}}=0$)
\cite{Goeke:2005hb,Metz:2008ib,Bacchetta-2007-JHEP}, 
so that the following sum rules should be satisfied:
\begin{align}
\int d^2 k_T \, h_{1LT}(x, k_T^{\, 2}) & = 0 ,
\nonumber \\[-0.05cm]
\int d^2 k_T \, g_{LT}(x, k_T^{\, 2}) & = 0 ,
\nonumber \\[-0.05cm]
\int d^2 k_T \,  h_{LL}(x, k_T^{\, 2}) & = 0 ,
\nonumber \\[-0.05cm]
\int d^2 k_T \, h_{3LT}(x, k_T^{\, 2}) & = 0 .
\label{eqn:cork-5}
\end{align} 
In the twist-2, although the collinear PDF $h_{1LT}(x)$ vanishes, 
its corresponding fragmentation function $H_{1LT}(z)$
exists as noted in Ref.\,\cite{Ji-1994}
as the function $\hat{h}_{\bar 1}$.
These T-odd terms are proportional to $(k_T)^0$ or $(k_T)^2$ 
in the correlation functions $\Phi^{[\Gamma]}$.
The terms with $(k_T)^1$ vanish and the term 
$k_T \cdot S_{TT} \cdot k_T$ also
vanishes after integrations, so there is no similar sum rule for other TMDs.
Similar sum rules exist for the TMDs $f_{1T}^\perp$ and $h$
in the spin-1/2 part as shown in Eqs.\,(22) and (23)
of Ref.\,\cite{Metz:2008ib}.
We may note that such a sum rule does not exist 
for the fragmentation functions
since the time-reversal invariance cannot be imposed
on the fragmentation functions, which contain 
the out-state $| P_h,\, S_h,\, X \rangle$
in its definition \cite{Ji-1994,bate2000,Goeke:2005hb,ffs-2016}.

\section{Summary}
\label{sec:4}

The possible TMDs were investigated for tensor-polarized spin-1 hadrons
by the  complete decomposition of the quark correlation function
including the lightcone vector $n$ in this work.
We found the 32 new terms which are dependent mainly on the vector $n$ 
in decomposing  the correlation function, so that 
there are totally 40 independent terms in the tensor-polarized correlation function. 
Furthermore, the tensor-polarized TMDs were studied up to twist-4 level
for the spin-1 hadron, and the 40 TMDs are found in association 
with the tensor polarization. 
There exist 10 TMDs in the twist-2 case.
Due to the existence of the new terms ($B_{20-52}$), 
the twist-2 TMD expressions 
of $f_{1LL}$, $f_{1LT}$, $g_{1LT}$, $h_{1LL}^\perp$, $h_{1LT}$, 
$h_{1LT}^\perp$ in terms of the expansion coefficients $A_i$
are modified. 
All the twist-3 and 4 TMDs (the following 30 TMDs) 
on the tensor-polarized spin-1 hadron:
\begin{itemize}
\vspace{-0.00cm}
\setlength{\leftskip}{2.80cm}
\setlength{\itemsep}{+0.00cm} 
\item[Twist-3 TMD:] $f_{LL}^\perp$, $e_{LL}$, 
      $f_{LT}$, $f_{LT}^\perp$, $e_{1T}$, $e_{1T}^\perp$,\\
      $f_{TT}$, $f_{TT}^\perp$, $e_{TT}$, $e_{TT}^\perp$,
      $g_{LL}^\perp$, $g_{LT}$, $g_{LT}^\perp$,\\
      $g_{TT}$, $g_{TT}^\perp$,
      $h_{1L}$, $h_{LT}$, $h_{LT}^\perp$, $h_{TT}$, $h_{TT}^\perp$,
\item[Twist-4 TMD:]  $f_{3LL}$, $f_{3LT}$, $f_{3TT}$, $g_{3LT}$, $g_{3TT}$, 
      $h_{3LL}^\perp$, $h_{3LT}$, $h_{3LT}^\perp$, $h_{3TT}$, $h_{3TT}^\perp$, 
\vspace{-0.00cm}
\end{itemize}
are new functions we found in this work.
We also found new sum rules for the TMDs as
$ \int d^2 k_T g_{LT} 
 = \int d^2 k_T h_{LL} = \int d^2 k_T h_{3LL} =0$.
Integrating these new TMDs, we found the collinear PDFs 
\begin{itemize}
\vspace{-0.10cm}
\setlength{\leftskip}{2.80cm}
\setlength{\itemsep}{-0.00cm} 
\item[Twist-3 PDF:] $e_{LL}$, $f_{LT}$, 
\item[Twist-4 PDF:] $f_{3LL}$, 
\vspace{-0.10cm}
\end{itemize}
in this work.
In addition, we explained that 
the corresponding transverse-momentum-dependent 
fragmentation functions exist for the tensor-polarized spin-1 hadrons.

Recently, the T-odd TMDs attract considerable attention 
since they are related to single spin asymmetries in the proton reactions. 
The T-odd TMDs in the spin-1 deuteron are also interesting 
to be investigated in future. 
Since there are projects to investigate the structure functions
of the polarized spin-1 deuteron at JLab, Fermilab, NICA,
and EIC, we hope that these new structure functions will be
experimentally investigated in future.

\begin{acknowledgements}

The authors thank A. Bacchetta for suggestions for understanding
Ref.\,\cite{bate2000}, and they thank P. J. Mulders
for usual comments.
The work of S.K. was partially supported by 
Japan Society for the Promotion of Science (JSPS) Grants-in-Aid 
for Scientific Research (KAKENHI) Grant Number 19K03830.
The work of Q.-T.S. was supported by the National Natural Science Foundation 
of China under Grant Number 12005191 and  the Academic Improvement Project 
of Zhengzhou University.
\end{acknowledgements}




\end{document}